\documentclass[prl,aps,twocolumn,superscriptaddress]{revtex4}

\usepackage{xcolor}
\usepackage{graphicx}
\usepackage[caption=false]{subfig}
\usepackage{amsmath}
\usepackage{booktabs}
\usepackage{rotating}
\usepackage{multirow}
\usepackage{bm}
\usepackage{txfonts}
\usepackage{hyperref}

\date{\today}
\newcommand{\be}{\begin{eqnarray}}
\newcommand{\ee}{\end{eqnarray}}
\newcommand{\bfb}{{\bf b}_{\perp}}
\newcommand{\Dp}{{\bf \Delta}_{\perp}}

\begin{document}

\title{Extraction of the Pion Unpolarized Quark Generalized Parton Distribution
at Zero Skewness from Charge Form Factors}

\author{Satyajit Puhan}
\email[]{puhansatyajit@gmail.com}
\affiliation{Computational High Energy Physics Lab, Department of Physics,
Dr.\ B.R.\ Ambedkar National Institute of Technology, Jalandhar, Punjab 144008, India}
\affiliation{Institute of Physics, Academia Sinica, Taipei 11529, Taiwan}

\author{Shubham Sharma}
\email[]{s.sharma.hep@gmail.com}
\affiliation{Laboratory for Advanced Scientific Technologies of Mega-Science Facilities and Experiments, Moscow Institute of Physics and Technology (MIPT), Dolgoprudny 141700, Russia}

\author{Narinder Kumar}
\email[]{narinderhep@gmail.com}
\affiliation{Computational Theoretical High Energy Physics Lab, Department of Physics, Doaba College, Jalandhar 144004, India}

\author{Harleen Dahiya}
\email[]{dahiyah@nitj.ac.in}
\affiliation{Computational High Energy Physics Lab, Department of Physics, Dr. B.R. Ambedkar National Institute of Technology, Jalandhar 144008, India}

\date{\today}

\begin{abstract}
Based on a fit to experimental and lattice-QCD measurements of the pion
electromagnetic form factor, using three pion parton distribution function
(PDF) inputs, JAM21, xFitter and GRV, we report a data-driven determination of
the unpolarized quark generalized parton distributions (GPDs) of the pion in
the zero-skewness limit ($\xi = 0$). The form factor is parameterized using a
flexible functional form constrained by data and embedded into a GPD framework
constructed from collinear PDFs and a profile function encoding transverse
dynamics. Three independent fits are reported, one to the experimental data
alone, one to the lattice-QCD data alone and one to the two together, which
allows the compatibility of the two sources of information to be assessed
directly. The uncertainty of the extracted GPD combines the $\Delta\chi^2=1$
region of the profile parameters with the propagated uncertainty of the input
PDF error members. We consider two GPD parameterizations, P-1 and P-2, to
study the sensitivity of the extracted GPDs to the choice of parameterization
and PDF input. This approach provides a unified description of the pion's
electromagnetic structure and its spatial parton distributions. We present the
extracted pion GPDs and their impact-parameter-space interpretations, offering
new insights into the internal structure of the lightest QCD bound state and
providing essential input for future studies at the Electron-Ion Collider via
the Sullivan process, exclusive $\pi^+$ electroproduction at the 12~GeV
Jefferson Lab program, pion-induced exclusive measurements at COMPASS,
proposed pion-beam experiments at AMBER, and phenomenological and lattice
investigations for structure of the meson.
\end{abstract}
\maketitle

{\it\textbf{1. Introduction:}}
Understanding the three-dimensional (3-D) structure of hadrons in terms of
their quark and gluon degrees of freedom remains a central challenge in
Quantum Chromodynamics (QCD). As the lightest QCD bound state and the pseudo-Goldstone boson associated with the spontaneous (dynamical) breaking of chiral symmetry, the pion occupies a
unique role in hadron structure studies. Its small but finite mass arises from
the explicit breaking of chiral symmetry by the nonzero light-quark masses.
Despite its apparent simplicity as a quark-antiquark system and its importance
in mediating the nucleon-nucleon interaction, the pion's 3-D partonic
structure remains far less constrained than that of the nucleon.\\

Generalized parton distributions (GPDs) provide
a unified framework that encodes both the longitudinal momentum and transverse
spatial distributions of partons, offering direct insight into the internal
dynamics of hadrons \cite{Diehl:2003ny,Belitsky:2005qn,Meissner:2008ay}. GPDs provide
access to a wide range of hadronic properties, including mechanical
distributions such as pressure, energy and forces, as well as fundamental
observables like charge radii, angular momentum and spatial densities
\cite{Chavez:2021llq}. The pion GPDs have been extensively investigated within
a variety of theoretical frameworks, including the Nambu--Jona-Lasinio (NJL)
model \cite{Zhang:2025xtn,Shastry:2022obb}, chiral quark models
\cite{Nematollahi:2024wrj,Fanelli:2016aqc}, Dyson--Schwinger equation
approaches \cite{Mezrag:2015mka,Mezrag:2014jka}, the Bethe--Salpeter
wave-function formalism \cite{Shi:2020pqe}, Basis Light-Front Quantization
(BLFQ) \cite{Adhikari:2021jrh}, continuum Schwinger-function analyses
\cite{Morgado-Chavez:2026hyt}, and lattice QCD simulations
\cite{Lin:2023gxz,Ding:2024saz}. These studies have provided valuable insights
into the pion's three-dimensional structure and have established important
theoretical benchmarks for pion GPDs.

Compared with the proton, phenomenological extractions of pion GPDs directly
constrained by experimental data remain rather limited because of the absence
of free pion targets
\cite{Diehl:2004cx,Cichy:2024afd,Hashamipour:2020kip,Diehl:2013xca,Guo:2025muf,Kriesten:2019jep}.
To date, only one dedicated data-driven extraction of the pion GPDs has been
reported by the MMGPDs group \cite{Goharipour:2025zsw}, employing the JAM21,
xFitter and MAP23 pion PDF sets
\cite{Barry:2021osv,Novikov:2020snp,Pasquini:2023aaf} together with a
Regge-inspired profile function. The pion GPDs can nevertheless be probed
through exclusive processes involving pion exchange or production, such as
exclusive electroproduction ($ep \rightarrow e^{\prime}\pi^{+}n$) at the
upgraded 12~GeV Jefferson Lab \cite{Arrington:2021alx}, pion-induced hard
exclusive reactions ($\pi^{-}p \rightarrow \gamma^{*}n$) at COMPASS and AMBER
\cite{Adams:2018pwt}, and the future tagged pion-exchange (Sullivan) process
at the electron-ion collider (EIC) \cite{Aguilar:2019teb}.

In the present work, we employ the Regge-inspired profile function proposed in
Ref.~\cite{Vaziri:2023xee} together with three pion PDF inputs, JAM21,
xFitter and GRV. The corresponding valence pion PDFs are used directly in the
GPD construction, without any intermediate refit, while the parameters of the
profile function are constrained through a fit to the available experimental
and lattice-QCD pion electromagnetic form-factor data. We consider two GPD
parameterizations, referred to as P-1 and P-2. The parameterization P-2 is the
Regge-inspired profile function previously used by the MMGPDs group in
Ref.~\cite{Goharipour:2025zsw}, and we include it here purely for comparison,
so that the sensitivity of the extracted GPDs to the choice of profile can be
assessed within an otherwise identical analysis. This provides an independent
determination of the pion GPDs and enables an assessment of the model
dependence associated with different pion PDF inputs and GPD
parameterizations.

The three inputs are constrained by different data. JAM21
\cite{Barry:2021osv} and xFitter \cite{Novikov:2020snp} are next-to-leading
order analyses of pion-induced Drell--Yan production together with
leading-neutron deep-inelastic scattering, delivered as replicas and as
symmetric Hessian members respectively, whereas the older GRV set
\cite{Gluck:1991ey} rests on Drell--Yan and prompt-photon data alone, predates
the leading-neutron measurements that constrain the small-$x$ valence
distribution, and provides no uncertainty estimate.

Precisely stating the scope of the present analysis in the manuscript, the
electromagnetic form factor is a single $x$-integrated observable and cannot
by itself resolve the $x$ dependence of $H^q(x,0,t)$. The present result is
accordingly a model-dependent determination of the GPD rather than a global
fit in the sense in which that term is used for PDF analyses. The quantity
constrained by the data is the $x$-$t$ correlation carried by the profile
function for a given functional form, and the sensitivity to that choice is
assessed by performing the complete analysis with both P-1 and P-2.
\begin{figure}[t]
  \centering
  \includegraphics[width=\columnwidth]{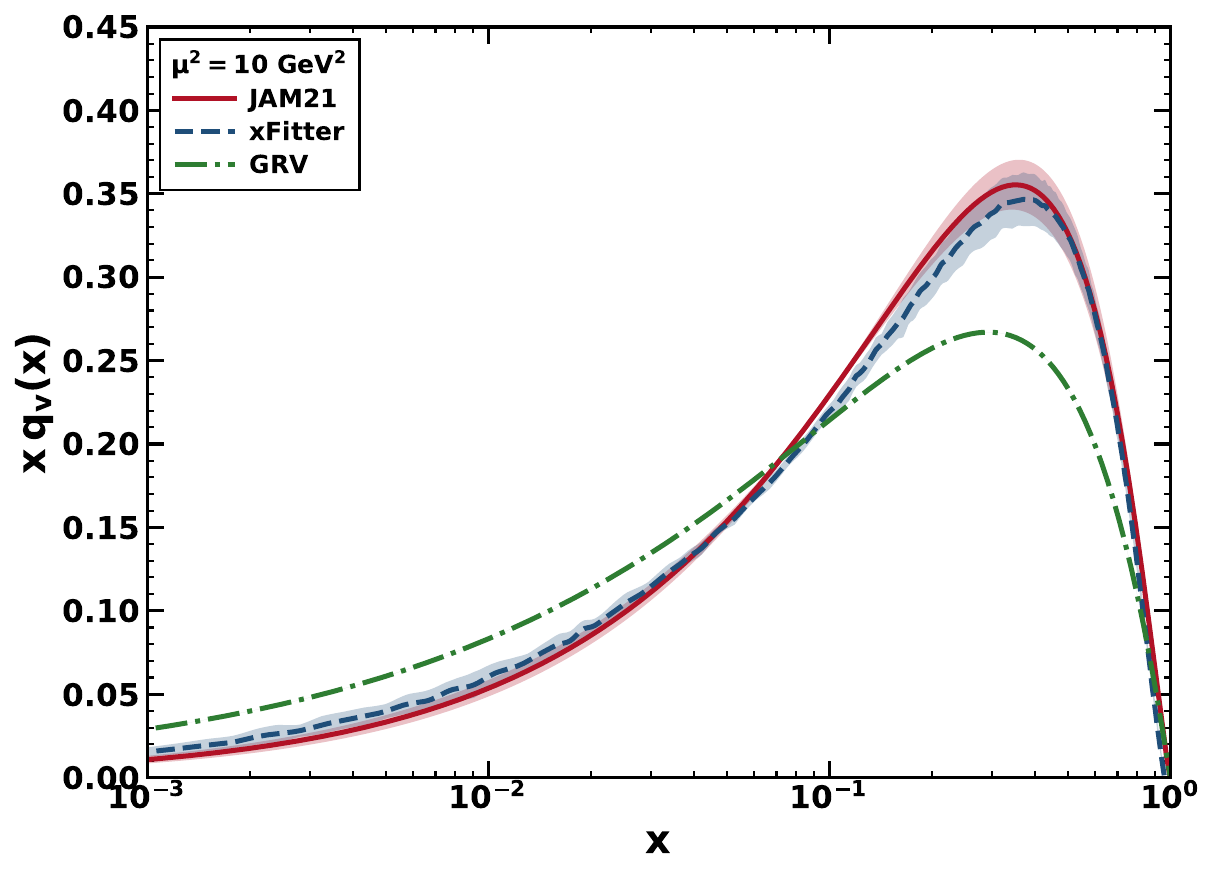}
  \caption{
  Valence-quark distributions $x q_v(x)$ of the pion used as input to the
  present analysis, shown at $\mu^2 = 10~\mathrm{GeV}^2$. The solid red,
  dashed blue and dash-dotted green curves are the JAM21
  \cite{Barry:2021osv}, xFitter \cite{Novikov:2020snp} and GRV
  \cite{Gluck:1991ey} sets, taken directly from the corresponding LHAPDF
  grids. The bands show the uncertainty of each set, obtained from its own
  error members. The GRV set provides a central member only and therefore
  carries no band.
  }
  \label{fig:pdfs}
\end{figure}

{\it\textbf{2. Phenomenological Framework.}}
For a pseudoscalar meson only a single chiral-even quark GPD contributes at
leading twist. We use the standard operator definition of the unpolarized
quark GPD \cite{Diehl:2003ny}, in which the light-cone correlator of the quark
vector current is decomposed as
\begin{align}
H^q(x,\xi,t) =
\frac{1}{2}\int \frac{dz^-}{2\pi}\, e^{\,i x P^{+} z^{-}}
\left\langle p^{\prime}\left|
\bar{q}\!\left(-\tfrac{z}{2}\right)\gamma^{+}
q\!\left(\tfrac{z}{2}\right)
\right| p \right\rangle
\bigg|_{z^{+}=0,\; \bm{z}_{\perp}=0},
\label{eq:def}
\end{align}
where $P = (p+p^{\prime})/2$, a light-cone gauge link between the quark
fields is understood, $t = (p^{\prime}-p)^2$ is the squared momentum transfer,
$\xi = (p^{+}-p^{\prime +})/(p^{+}+p^{\prime +})$ is the skewness and $x$ is
the average light-cone momentum fraction carried by the struck quark. With this normalization the
forward limit of the GPD is the leading-twist quark distribution itself,
\begin{equation}
H^q(x,0,0) = q(x), 
\label{eq:forward}
\end{equation}
and $H^q(-x,0,0) = -\bar{q}(x)$ for the antiquark sector. This is the
convention that is consistent with the first-moment sum rule for the
electromagnetic form factor, which reads
\begin{equation}
F(t) = \sum_q e_q \int_{-1}^{1} dx\, H^q(x,0,t),
\label{eq:sumrule}
\end{equation}
and which is independent of $\xi$ by virtue of the polynomiality property of
GPDs. Throughout this work we quote the momentum transfer as $|t| = -t$ when
we refer to a range or to an axis label, and as $t$, which is negative in the
physical region, when we refer to a value of the argument of the GPD. The fit
covers $0.0138 \le |t| \le 9.77~\mathrm{GeV}^2$.

For the $\pi^{+}$, which has valence content $u\bar{d}$, charge conjugation
and isospin symmetry relate the two valence distributions,
\begin{equation}
q_v(x) \;\equiv\; u^{\pi^{+}}(x) - \bar{u}^{\pi^{+}}(x)
\;=\; \bar{d}^{\pi^{+}}(x) - d^{\pi^{+}}(x),
\label{eq:chargesym}
\end{equation}
so that a single independent valence distribution specifies the pion GPD.
Inserting Eq.~(\ref{eq:chargesym}) into Eq.~(\ref{eq:sumrule}) and using
$e_u = 2/3$ and $e_{\bar d} = 1/3$ gives
\begin{equation}
F(t) = \int_{0}^{1} dx\, H(x,0,t),
\label{eq:Fsimple}
\end{equation}
with the valence number sum rule $\int_0^1 dx\, q_v(x) = 1$ guarantee the
charge normalization $F(0)=1$ exactly, independently of any parameter of the
$t$-dependent profile. Equation~(\ref{eq:forward}) contains no additional
factor of $x$. 

At zero skewness, GPD can be defined in the factorized form
\begin{equation}
H(x,0,t) = q_v(x)\,\exp\!\left[\,\mathcal{S}(x,t)\,\right],
\label{eq:ansatz}
\end{equation}
where $q_v(x)$ is taken directly from a global pion PDF extraction and
$\mathcal{S}(x,t)$ is the profile function, which vanishes at $t=0$ and
carries the entire $x$-$t$ correlation. The exponential form guarantees
positivity of $H$ and reproduces the Regge behaviour
$H \sim x^{-\alpha^{\prime}t} q_v(x)$ at small $x$ when
$\mathcal{S} \rightarrow -\alpha^{\prime} t \ln x$. The first of the two profiles
considered here, P-1, is
\begin{equation}
\mathcal{S}_{\rm P\text{-}1}(x,t) =
-\alpha^{\prime}\,t\,(1-x)^{\gamma}\ln x
+ \beta\, x^{m} \ln\!\left(1 - b\, t\right),
\label{eq:P1}
\end{equation}
with five parameters $\alpha^{\prime}$, $\gamma$, $\beta$, $m$ and $b$. The
first term is the Regge term, with $(1-x)^{\gamma}$ switching off the
diffusion of the transverse size as $x \rightarrow 1$. The second term is
qualitatively different from the profiles used previously, because
$\ln(1-bt)$ grows only logarithmically with $|t|$, so that $H$ falls as a
power of $t$ rather than exponentially at fixed $x$.

For comparison we also use the Regge-inspired profile employed by the MMGPDs
group in Ref.~\cite{Goharipour:2025zsw}, which we refer to as P-2,
\begin{equation}
\mathcal{S}_{\rm P\text{-}2}(x,t) =
t\left[\tilde{\alpha}^{\prime}(1-x)^{3}\ln\frac{1}{x}
+ B(1-x)^{3} + A\,x(1-x)^{2}\right],
\label{eq:P2}
\end{equation}
with three parameters $\tilde{\alpha}^{\prime}$, $B$ and $A$, the slope
$\tilde{\alpha}^{\prime}$ being distinguished from the $\alpha^{\prime}$ of
Eq.~(\ref{eq:P1}) because the two multiply different functions of $x$. This form is linear in
$t$ in the exponent, so that $H$ falls exponentially with $|t|$ at fixed $x$,
in contrast with P-1. It is not proposed here as a new parameterization. The
two forms are fitted under identical conditions, with the same data, the same
PDF inputs and the same treatment of uncertainties, so that the model
dependence of the extracted $x$-$t$ correlation can be isolated.

The three PDF inputs, JAM21 \cite{Barry:2021osv}, xFitter
\cite{Novikov:2020snp} and GRV \cite{Gluck:1991ey}, are shown in
Fig.~\ref{fig:pdfs}. They are read from the corresponding LHAPDF grids at the
scale $\mu^2 = 10~\mathrm{GeV}^2$ at which the results of Sec.~3 are
quoted, and are used without any intermediate refit. This choice follows
directly from the observation that fitting a new functional form to an
existing extraction discards the correlated uncertainty of that extraction
while adding nothing that is constrained by data. Working with the released
sets instead allows the quoted PDF uncertainty to be propagated member by
member, as described below.

The profile parameters are determined by minimizing
\begin{align}
\chi^2 =
\sum_{i \in F}
\left[
\frac{F(t_i) - F_i^{\rm exp}}{\sigma_i}
\right]^2
+
\sum_{j \in |F|^2}
\left[
\frac{|F(t_j)|^2 - (|F|^2)_j^{\rm exp}}{\sigma_j}
\right]^2 ,
\label{eq:chi2}
\end{align}
where the first sum runs over the data quoted as $F(t)$, the second over the
data quoted as $|F(t)|^2$, and $\sigma_i$ is the total uncertainty of the
corresponding point obtained by adding the statistical and systematic
uncertainties in quadrature as published. Correlated systematic and
normalization uncertainties are not released in a usable covariance form for
the majority of the data sets entering this analysis, so Eq.~(\ref{eq:chi2})
is diagonal by necessity. We regard this as a limitation rather than as a
choice and we state it explicitly, because the reduced $\chi^2$ values of the
individual sets in Table~\ref{tab:chi2} differ considerably and part of that
spread would very likely be absorbed by a proper treatment of normalization
uncertainties. No dataset-dependent weights are introduced. All parameters of both profiles,
including the Regge slope, are free and are determined by the data. The
minimization uses the CERN MINUIT implementation \cite{James:1975dr} with
{\tt errordef}$=1$, and is repeated from several starting points to guard
against local minima. The same procedure is applied to P-1 and to P-2 without
any change.

Three fits are performed and reported separately. Fit~E uses the experimental
form-factor data only, comprising the $F(t)$ measurements of
Refs.~\cite{Brown:1973wr,Ackermann:1977rp,Brauel:1979zk,JeffersonLabFpi:2000nlc,JeffersonLabFpi-2:2006ysh,JeffersonLabFpi:2007vir,JeffersonLab:2008jve,Bebek:1977pe}
and the $|F(t)|^2$ measurements of
Refs.~\cite{Adylov:1977kj,Dally:1981ur,Dally:1982zk,NA7:1986vav}, for a total
of $N=148$ points. Fit~L uses the lattice-QCD form factor of
Ref.~\cite{Ding:2024lfj} only, $N=13$. Fit~T uses the two together, $N=161$.

The uncertainty of every quantity derived from the fit has two independent
components which are added in quadrature. The first is the uncertainty of the
profile parameters at fixed PDF. Several parameters sit on or close to the
boundary of their allowed range, where the parabolic approximation to the
likelihood underlying the Hessian estimate breaks down, so that the covariance
matrix is not a faithful description of the likelihood and a linear
propagation of it underestimates the band
\cite{Cowan:2010js,James:1975dr}. We therefore construct the
interval directly from the likelihood, sampling it with a parallel-chain
Metropolis algorithm \cite{Metropolis:1953am,Hastings:1970aa} using 96 chains
with an adaptive step size, retaining every sampled point with
$\chi^2 \le \chi^2_{\rm min}+1$, recomputing the observable for each retained
point and taking the envelope of the resulting curves. This is the standard
$\Delta\chi^2 = 1$ prescription for a one-standard-deviation interval
\cite{Cowan:2010js,ParticleDataGroup:2022pth}, implemented by sampling rather than by
expansion about the minimum, and it therefore remains valid at a boundary,
where the Hessian estimate does not. Monte Carlo sampling of the likelihood is
by now standard practice in global analyses of hadron structure
\cite{Sato:2016tuz,Barry:2021osv}. The second component is the uncertainty of the input PDF.
The complete fit is repeated for the error members of the input set, and the
spread of the resulting observables is combined with the prescription
appropriate to that set, the replica standard deviation for JAM21 and the
symmetric Hessian sum for xFitter. The GRV set has no error members, so for
that input only the first component contributes and its bands are
correspondingly narrower. No part of the quoted uncertainty comes from
variation of an evolution scale.

\begin{figure*}[t]
  \centering
  \includegraphics[width=\textwidth]{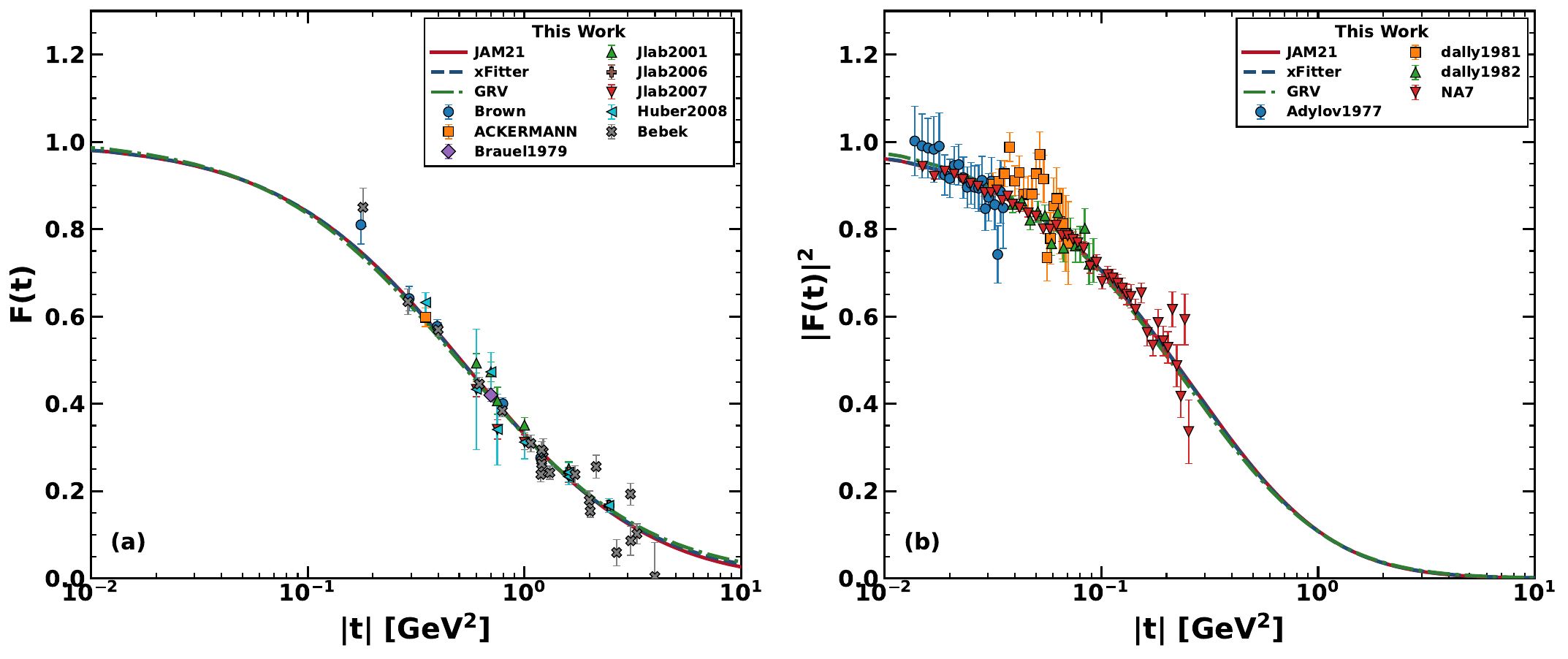}
  \caption{
  ({\bf Fit-E obtained from the experimental and lattice-QCD data together}) Panel (a)
  shows the pion electromagnetic form factor $F(t)$ and panel (b) the squared
  form factor $|F(t)|^2$, both as functions of $|t|$. The solid red, dashed
  blue and dash-dotted green curves are obtained with the JAM21, xFitter and
  GRV valence distributions respectively. The $F(t)$ data are taken from
  Refs.~\cite{Brown:1973wr,Ackermann:1977rp,Brauel:1979zk,JeffersonLabFpi:2000nlc,JeffersonLabFpi-2:2006ysh,JeffersonLabFpi:2007vir,JeffersonLab:2008jve,Bebek:1977pe}
  and the $|F(t)|^2$ data from
  Refs.~\cite{Adylov:1977kj,Dally:1981ur,Dally:1982zk,NA7:1986vav}. The
  goodness of fit for the individual sets is given in Table~\ref{tab:chi2}.
  }
  \label{fig:fitE}
\end{figure*}

\begin{figure}[t]
  \centering
  \includegraphics[width=\columnwidth]{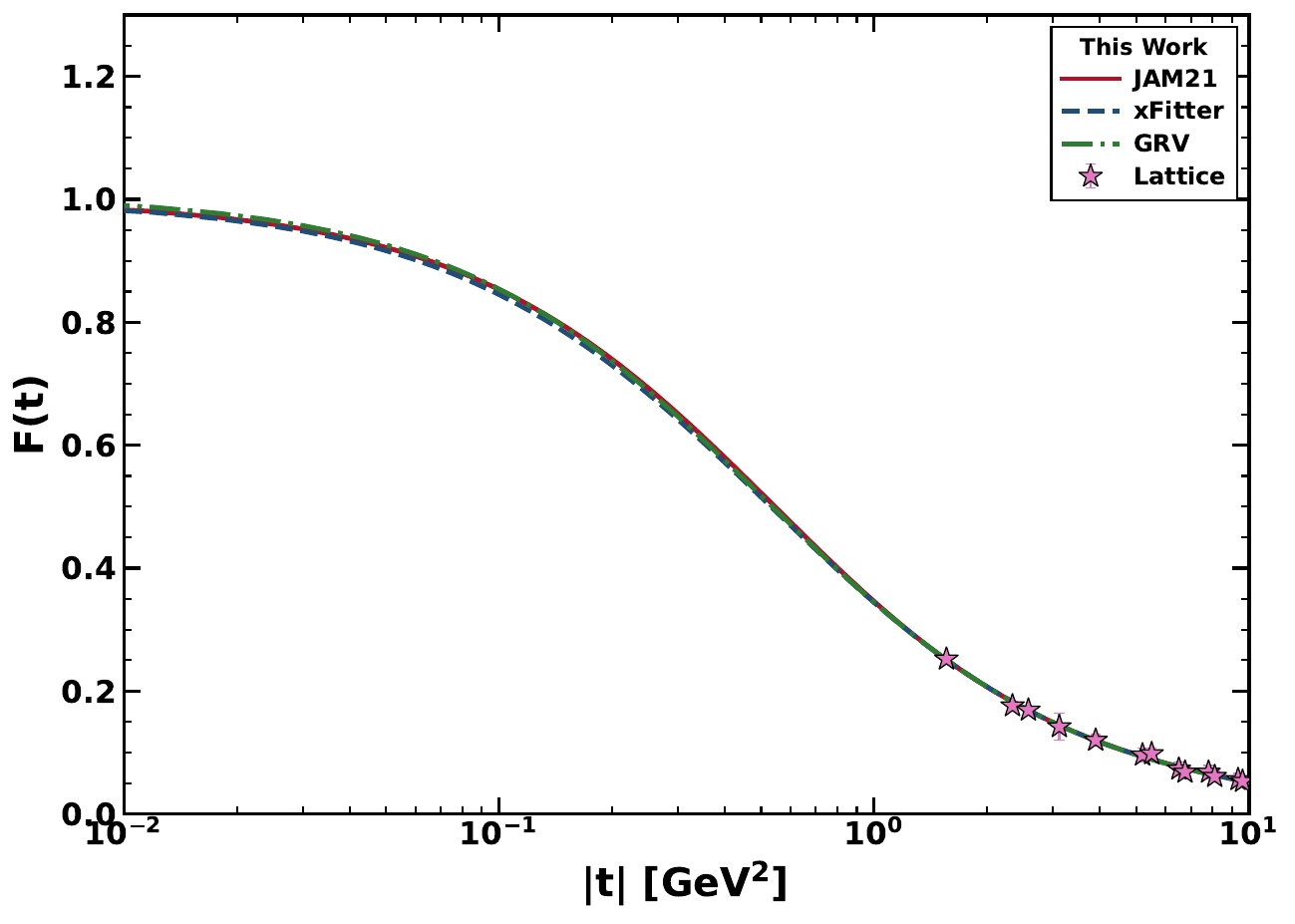}
  \caption{
  ({\bf Fit~L obtained from the lattice-QCD form factor of
  Ref.~\cite{Ding:2024lfj} alone}.) The curves follow the same convention as in
  Fig.~\ref{fig:fitE}. The lattice calculation provides $F(t)$ only, so there
  is no $|F(t)|^2$ counterpart to this figure.
  }
  \label{fig:fitL}
\end{figure}

\begin{figure*}[t]
  \centering
  \includegraphics[width=\textwidth]{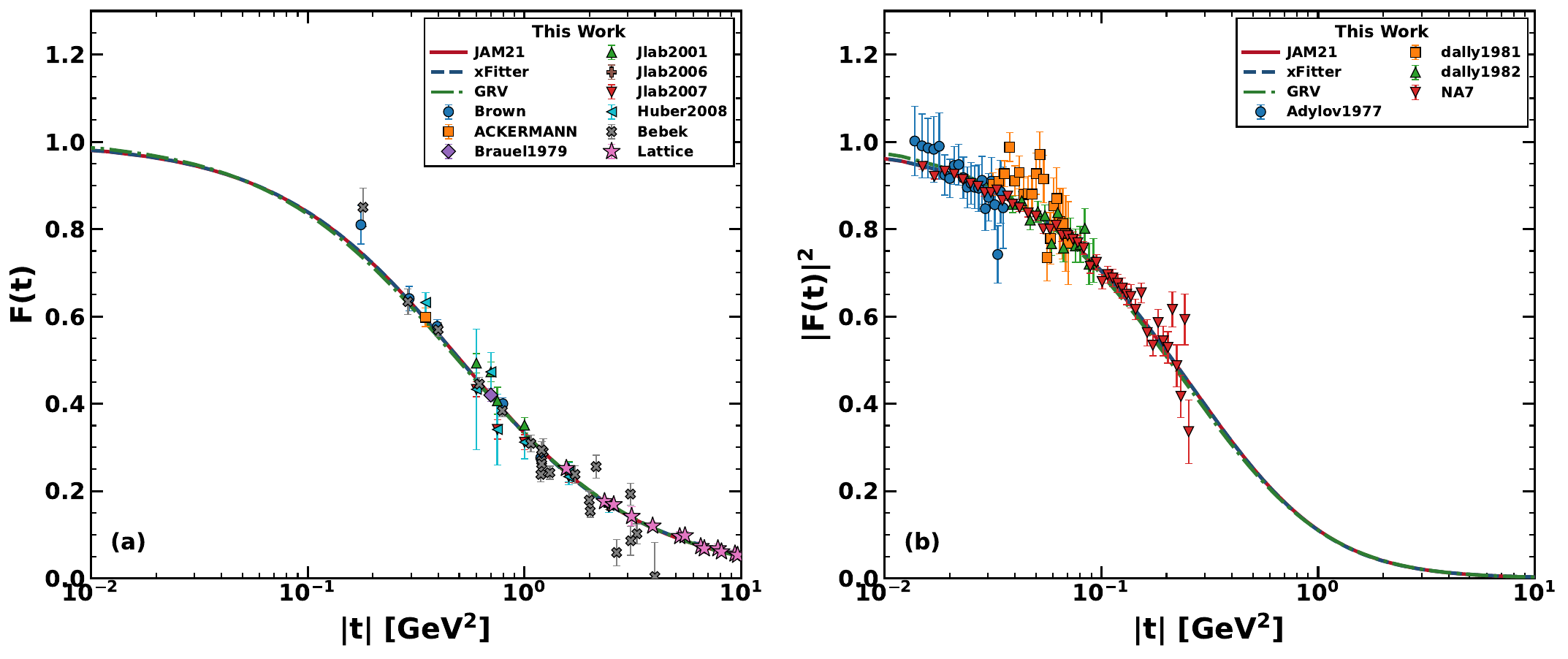}
  \caption{
  ({\bf Fit T obtained from the experimental and lattice-QCD data together}) Panel
  (a) shows $F(t)$, where the lattice points of Ref.~\cite{Ding:2024lfj} are
  included alongside the experimental measurements, and panel (b) shows
  $|F(t)|^2$. The curves and the data references are as in
  Fig.~\ref{fig:fitE}.
  }
  \label{fig:fitT}
\end{figure*}

\begin{table*}[t]
\caption{
Reduced $\chi^2$ per data set for the three fits and the three pion PDF
inputs. Fit~E uses the experimental data only, fit~L the lattice-QCD form
factor of Ref.~\cite{Ding:2024lfj} only, and fit~T the two together. A dash
indicates that the set does not enter the corresponding fit. The column $N$
gives the number of points in each set, and the last row gives the total
$\chi^2/N$.
}
\label{tab:chi2}
\centering
\begin{ruledtabular}
\begin{tabular}{lcccccccc}
 & & & \multicolumn{3}{c}{Fit E (experiment)}
   & \multicolumn{3}{c}{Fit L (lattice)} \\
Data set & Obs. & $N$ & JAM21 & xFitter & GRV & JAM21 & xFitter & GRV \\
\hline
Brown \cite{Brown:1973wr}                    & $F$     &   5 & 0.97 & 1.06 & 1.50 & -- & -- & -- \\
Ackermann \cite{Ackermann:1977rp}            & $F$     &   1 & 0.01 & 0.04 & 0.31 & -- & -- & -- \\
Brauel \cite{Brauel:1979zk}                  & $F$     &   1 & 0.02 & 0.08 & 0.24 & -- & -- & -- \\
JLab 2001 \cite{JeffersonLabFpi:2000nlc}     & $F$     &   5 & 2.39 & 2.59 & 2.80 & -- & -- & -- \\
JLab 2006 \cite{JeffersonLabFpi-2:2006ysh}   & $F$     &   2 & 1.65 & 1.43 & 0.87 & -- & -- & -- \\
JLab 2007 \cite{JeffersonLabFpi:2007vir}     & $F$     &   4 & 2.63 & 2.38 & 2.06 & -- & -- & -- \\
Huber 2008 \cite{JeffersonLab:2008jve}       & $F$     &   8 & 0.76 & 0.78 & 0.88 & -- & -- & -- \\
Bebek \cite{Bebek:1977pe}                    & $F$     &  21 & 2.83 & 2.80 & 2.96 & -- & -- & -- \\
Adylov \cite{Adylov:1977kj}                  & $|F|^2$ &  22 & 0.43 & 0.43 & 0.44 & -- & -- & -- \\
Dally 1981 \cite{Dally:1981ur}               & $|F|^2$ &  20 & 2.21 & 2.14 & 2.06 & -- & -- & -- \\
Dally 1982 \cite{Dally:1982zk}               & $|F|^2$ &  14 & 0.65 & 0.65 & 0.70 & -- & -- & -- \\
NA7 \cite{NA7:1986vav}                       & $|F|^2$ &  45 & 1.23 & 1.32 & 2.16 & -- & -- & -- \\
Lattice \cite{Ding:2024lfj}                  & $F$     &  13 &  --  &  --  &  --  & 0.18 & 0.19 & 0.19 \\
\hline
Total $\chi^2/N$ & & 148 / 13 & 1.45 & 1.46 & 1.75 & 0.18 & 0.19 & 0.19 \\
\end{tabular}

\vspace{6pt}

\begin{tabular}{lccccc}
 & & & \multicolumn{3}{c}{Fit T (experiment $+$ lattice)} \\
Data set & Obs. & $N$ & JAM21 & xFitter & GRV \\
\hline
Brown \cite{Brown:1973wr}                    & $F$     &   5 & 1.19 & 1.17 & 1.63 \\
Ackermann \cite{Ackermann:1977rp}            & $F$     &   1 & 0.05 & 0.05 & 0.36 \\
Brauel \cite{Brauel:1979zk}                  & $F$     &   1 & 0.02 & 0.04 & 0.14 \\
JLab 2001 \cite{JeffersonLabFpi:2000nlc}     & $F$     &   5 & 2.05 & 2.16 & 2.36 \\
JLab 2006 \cite{JeffersonLabFpi-2:2006ysh}   & $F$     &   2 & 0.15 & 0.19 & 0.16 \\
JLab 2007 \cite{JeffersonLabFpi:2007vir}     & $F$     &   4 & 2.73 & 2.60 & 2.36 \\
Huber 2008 \cite{JeffersonLab:2008jve}       & $F$     &   8 & 0.68 & 0.68 & 0.86 \\
Bebek \cite{Bebek:1977pe}                    & $F$     &  21 & 3.45 & 3.33 & 3.56 \\
Adylov \cite{Adylov:1977kj}                  & $|F|^2$ &  22 & 0.43 & 0.43 & 0.43 \\
Dally 1981 \cite{Dally:1981ur}               & $|F|^2$ &  20 & 2.19 & 2.16 & 2.09 \\
Dally 1982 \cite{Dally:1982zk}               & $|F|^2$ &  14 & 0.66 & 0.66 & 0.70 \\
NA7 \cite{NA7:1986vav}                       & $|F|^2$ &  45 & 1.28 & 1.31 & 2.15 \\
Lattice \cite{Ding:2024lfj}                  & $F$     &  13 & 0.62 & 0.69 & 0.52 \\
\hline
Total $\chi^2/N$ & & 161 & 1.45 & 1.44 & 1.72 \\
\end{tabular}
\end{ruledtabular}
\end{table*}

\begin{table*}[t]
\caption{
Best-fit parameters of the two profile functions for the three fits and the
three pion PDF inputs. The quoted intervals are the projections of the
$\Delta\chi^2 = 1$ region, combined in quadrature with the spread over the PDF
error members.
}
\label{tab:params}
\centering
\setlength{\tabcolsep}{3pt}\small
\begin{ruledtabular}
\begin{tabular}{llccccc}
Fit & PDF & $\alpha^{\prime}~[\mathrm{GeV}^{-2}]$ & $\gamma$ & $\beta$ & $m$ & $b~[\mathrm{GeV}^{-2}]$ \\
\hline
\multirow{3}{*}{E}
 & JAM21    & $0.915^{+0.045}_{-0.040}$ & $1.951^{+1.421}_{-1.331}$ & $-2.813^{+2.041}_{-1.893}$ & $5.000^{+0.000}_{-3.216}$ & $5.000^{+0.000}_{-2.656}$ \\
 & xFitter  & $0.962^{+0.038}_{-0.046}$ & $0.409^{+1.860}_{-0.309}$ & $0.364^{+0.611}_{-1.793}$ & $0.309^{+1.828}_{-0.209}$ & $1.159^{+1.187}_{-0.982}$ \\
 & GRV      & $0.987^{+0.035}_{-0.032}$ & $0.100^{+0.057}_{-0.000}$ & $0.578^{+0.028}_{-0.048}$ & $0.174^{+0.081}_{-0.067}$ & $2.081^{+0.660}_{-0.442}$ \\
\hline
\multirow{3}{*}{L}
 & JAM21    & $0.949^{+0.047}_{-0.045}$ & $0.101^{+2.327}_{-0.001}$ & $0.623^{+0.199}_{-1.019}$ & $1.846^{+1.836}_{-1.491}$ & $5.000^{+0.000}_{-2.317}$ \\
 & xFitter  & $0.949^{+0.041}_{-0.048}$ & $0.101^{+1.461}_{-0.001}$ & $1.571^{+0.209}_{-2.325}$ & $0.100^{+1.862}_{-0.000}$ & $0.274^{+1.631}_{-0.264}$ \\
 & GRV      & $0.950^{+0.045}_{-0.046}$ & $0.711^{+0.081}_{-0.594}$ & $0.231^{+0.333}_{-0.028}$ & $0.100^{+0.660}_{-0.000}$ & $5.000^{+0.000}_{-1.654}$ \\
\hline
\multirow{3}{*}{T}
 & JAM21    & $0.939^{+0.051}_{-0.049}$ & $1.613^{+0.851}_{-0.674}$ & $-0.089^{+0.458}_{-0.479}$ & $0.431^{+1.361}_{-0.331}$ & $4.999^{+0.001}_{-2.602}$ \\
 & xFitter  & $0.953^{+0.027}_{-0.044}$ & $0.100^{+0.926}_{-0.000}$ & $1.793^{+0.378}_{-1.320}$ & $0.195^{+0.875}_{-0.095}$ & $0.236^{+0.270}_{-0.222}$ \\
 & GRV      & $0.967^{+0.034}_{-0.044}$ & $0.349^{+0.181}_{-0.190}$ & $0.507^{+0.182}_{-0.129}$ & $0.105^{+0.148}_{-0.005}$ & $1.710^{+0.982}_{-0.485}$ \\
\end{tabular}

\vspace{8pt}

\begin{tabular}{llccc}
Fit & PDF & $\tilde{\alpha}^{\prime}~[\mathrm{GeV}^{-2}]$ & $B$ & $A$ \\
\hline
\multirow{3}{*}{E}
 & JAM21    & $0.961^{+0.046}_{-0.047}$ & $0.329^{+0.216}_{-0.214}$ & $1.705^{+0.273}_{-0.270}$ \\
 & xFitter  & $0.948^{+0.044}_{-0.047}$ & $0.074^{+0.260}_{-0.256}$ & $1.997^{+0.356}_{-0.354}$ \\
 & GRV      & $0.986^{+0.041}_{-0.043}$ & $-0.794^{+0.184}_{-0.182}$ & $2.596^{+0.339}_{-0.329}$ \\
\hline
\multirow{3}{*}{L}
 & JAM21    & $0.948^{+0.049}_{-0.049}$ & $0.577^{+0.197}_{-0.193}$ & $0.856^{+0.186}_{-0.186}$ \\
 & xFitter  & $0.947^{+0.048}_{-0.045}$ & $0.497^{+0.206}_{-0.200}$ & $0.776^{+0.194}_{-0.185}$ \\
 & GRV      & $0.945^{+0.049}_{-0.048}$ & $-0.274^{+0.136}_{-0.134}$ & $1.094^{+0.172}_{-0.170}$ \\
\hline
\multirow{3}{*}{T}
 & JAM21    & $0.894^{+0.043}_{-0.042}$ & $0.698^{+0.171}_{-0.173}$ & $0.908^{+0.147}_{-0.144}$ \\
 & xFitter  & $0.844^{+0.045}_{-0.046}$ & $0.613^{+0.174}_{-0.173}$ & $0.909^{+0.172}_{-0.166}$ \\
 & GRV      & $0.861^{+0.032}_{-0.031}$ & $-0.149^{+0.113}_{-0.118}$ & $1.179^{+0.159}_{-0.153}$ \\
\end{tabular}
\end{ruledtabular}
\vspace{2pt}
\raggedright
\footnotesize The upper block gives the five parameters of P-1 and the lower
block the three parameters of P-2, Eq.~(\ref{eq:P2}).
\end{table*}

\begin{table*}[t]
\caption{
Pion charge radius $r_\pi$ and valence-averaged impact-parameter width
$\langle b_\perp^2\rangle$ obtained from the fits of Table~\ref{tab:params},
for both profile functions. The uncertainties are the total defined in
Sec.~2, that is the $\Delta\chi^2 = 1$ interval combined in quadrature with
the propagated PDF uncertainty.
}
\label{tab:obs}
\centering
\begin{ruledtabular}
\begin{tabular}{llcccc}
 & & \multicolumn{2}{c}{P-1} & \multicolumn{2}{c}{P-2} \\
Fit & PDF & $r_\pi~[\mathrm{fm}]$ & $\langle b_\perp^2\rangle~[\mathrm{fm}^2]$ & $r_\pi~[\mathrm{fm}]$ & $\langle b_\perp^2\rangle~[\mathrm{fm}^2]$ \\
\hline
\multirow{3}{*}{E}
 & JAM21    & $0.6697 \pm 0.0063$ & $0.2990 \pm 0.0057$ & $0.6619 \pm 0.0039$ & $0.2921 \pm 0.0034$ \\
 & xFitter  & $0.6630 \pm 0.0042$ & $0.2931 \pm 0.0037$ & $0.6648 \pm 0.0041$ & $0.2947 \pm 0.0036$ \\
 & GRV      & $0.6956 \pm 0.0033$ & $0.3226 \pm 0.0030$ & $0.6971 \pm 0.0040$ & $0.3239 \pm 0.0037$ \\
\hline
\multirow{3}{*}{L}
 & JAM21    & $0.6251 \pm 0.0390$ & $0.2605 \pm 0.0335$ & $0.6690 \pm 0.0124$ & $0.2984 \pm 0.0112$ \\
 & xFitter  & $0.6508 \pm 0.0252$ & $0.2824 \pm 0.0222$ & $0.6879 \pm 0.0146$ & $0.3154 \pm 0.0135$ \\
 & GRV      & $0.6364 \pm 0.0690$ & $0.2700 \pm 0.0617$ & $0.7180 \pm 0.0129$ & $0.3437 \pm 0.0123$ \\
\hline
\multirow{3}{*}{T}
 & JAM21    & $0.6665 \pm 0.0046$ & $0.2961 \pm 0.0040$ & $0.6663 \pm 0.0037$ & $0.2959 \pm 0.0033$ \\
 & xFitter  & $0.6638 \pm 0.0039$ & $0.2937 \pm 0.0034$ & $0.6696 \pm 0.0039$ & $0.2989 \pm 0.0035$ \\
 & GRV      & $0.6969 \pm 0.0042$ & $0.3238 \pm 0.0039$ & $0.6984 \pm 0.0039$ & $0.3252 \pm 0.0036$ \\
\end{tabular}
\end{ruledtabular}
\end{table*}

{\it\textbf{3. Results and Discussion.}}
The three fits are shown in Figs.~\ref{fig:fitE}, \ref{fig:fitL} and
\ref{fig:fitT}, the goodness of fit is collected in Table~\ref{tab:chi2}, and
the resulting parameters are collected in Table~\ref{tab:params} and the
derived observables in Table~\ref{tab:obs}. We discuss in turn what is
constrained by the data, what is inherited from the PDF input and what follows
from the assumed parameterization.

Fit~E gives $\chi^2/N = 1.45$, $1.46$ and $1.75$ for JAM21, xFitter and GRV,
and fit~T gives $1.45$, $1.44$ and $1.72$ for the same three inputs. The two
sets of numbers are essentially identical, which is the first and most direct
statement that can be made about compatibility. Adding thirteen lattice points
that populate the region $|t| \gtrsim 1.5~\mathrm{GeV}^2$ does not degrade the
description of the $148$ experimental points at all, and the lattice points
themselves are described with $\chi^2/N$ between $0.52$ and $0.69$ in the
combined fit, compared with $0.18$ to $0.19$ when they are fitted alone. In
other words, the lattice data are pulled only mildly by the experimental data,
and the two are consistent within the quoted uncertainties over the range
where they overlap. The comparison should nevertheless be read with the caveat that
the lattice uncertainties are large relative to the point-to-point scatter,
which is why fit~L reaches $\chi^2/N \simeq 0.19$. A data set that is fitted
with a reduced $\chi^2$ well below unity does not discriminate strongly
between models, so the agreement between fit~E and fit~T is evidence of the
absence of tension rather than evidence of a stringent cross-check.

Among the individual sets, the largest values are found for the
electroproduction data of Bebek {\it et al.}~\cite{Bebek:1977pe},
$\chi^2/N \simeq 3.4$, for the JLab 2001 and 2007 sets, $\chi^2/N \simeq 2.1$
to $2.7$, and for Dally 1981 \cite{Dally:1981ur}, $\chi^2/N \simeq 2.2$. These
values are stable across the three PDF inputs and across the three fits, which
indicates that they reflect tension among the measurements themselves rather
than a deficiency of the profile function. As noted above, the absence of
published correlated normalization uncertainties for these sets prevents a
quantitative test of that statement.

The fitted Regge slope is worth noting in its own right. Across all nine fits
and both profiles it comes out between $0.84$ and $0.99~\mathrm{GeV}^{-2}$,
with $\alpha^{\prime} = 0.939$, $0.953$ and $0.967~\mathrm{GeV}^{-2}$ for P-1
in the combined fit with the JAM21, xFitter and GRV inputs, and
$\tilde{\alpha}^{\prime} = 0.894$, $0.844$ and $0.861~\mathrm{GeV}^{-2}$ for
P-2. These values were not imposed. They emerge from the form-factor data
alone and sit close to the slope of the $\rho$ trajectory, which supports the
Regge interpretation of the small-$x$ behaviour built into
Eqs.~(\ref{eq:P1}) and (\ref{eq:P2}) and indicates that the profile is
behaving as intended in the region where the Regge term dominates.

The GRV input is clearly disfavoured relative to the other two, by
$\Delta\chi^2 \simeq 44$ over $161$ points in the combined fit. The reason is
visible in Fig.~\ref{fig:pdfs}. The GRV valence distribution is harder at small
$x$ and softer at large $x$ than JAM21 and xFitter, and since $F(t)$ is the
$x$ integral of $q_v(x)\exp[\mathcal{S}]$, that shape cannot be compensated by
the profile without distorting the low-$|t|$ behaviour, which controls the
charge radius. Consistently, GRV gives $r_\pi = 0.6969 \pm 0.0042~\mathrm{fm}$,
about eight standard deviations above the world average.

\begin{figure}[t]
  \centering
  \includegraphics[width=\columnwidth]{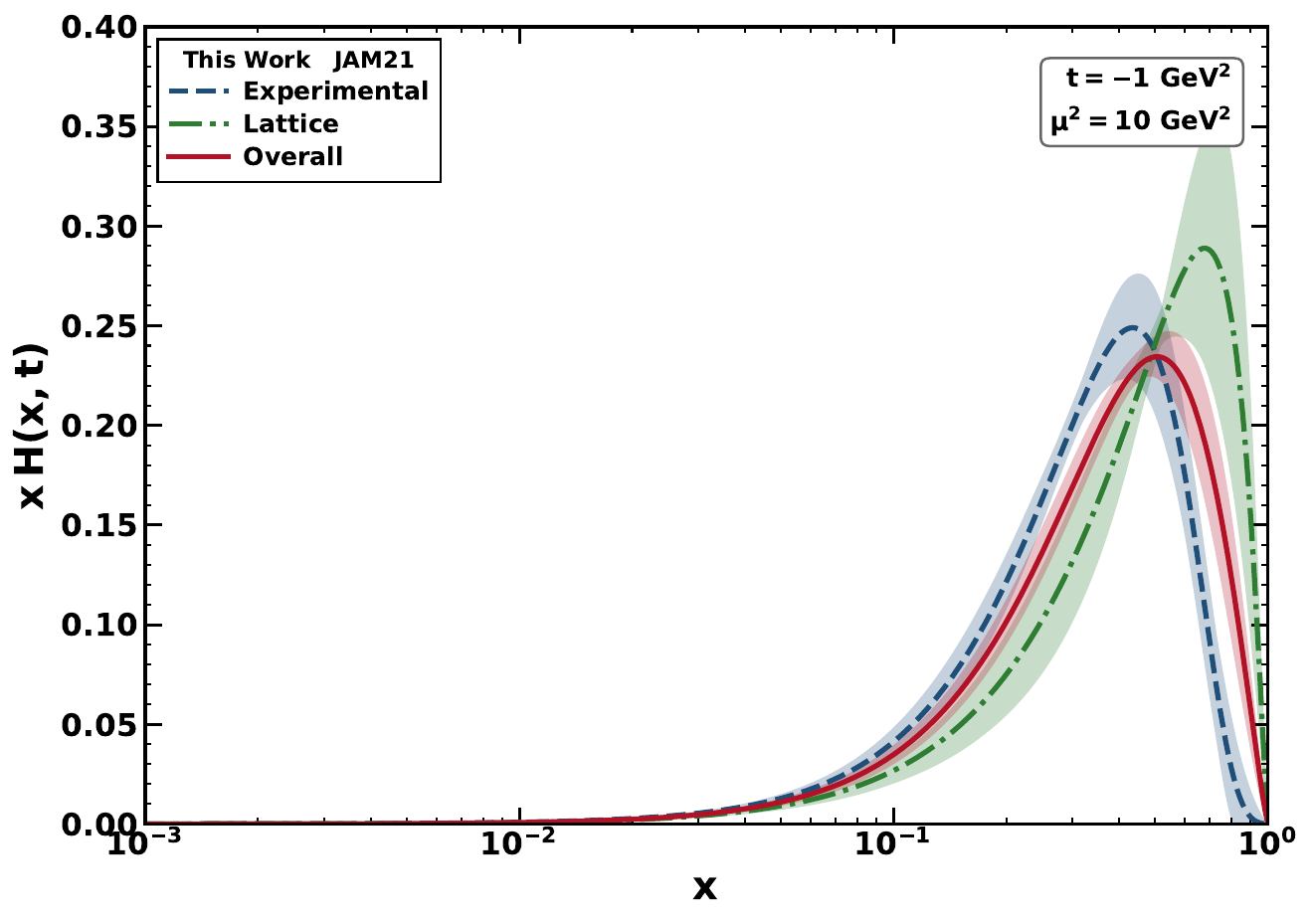}
  \caption{
  Extracted GPD, shown as $x H(x,0,t)$ at fixed $t = -1~\mathrm{GeV}^2$ and
  $\mu^2 = 10~\mathrm{GeV}^2$ with the JAM21 valence distribution, for the
  three fits. The dashed blue curve is fit~E, the dash-dotted green curve is
  fit~L and the solid red curve is fit~T. The shaded bands are the total
  uncertainties defined in Sec.~2.
  }
  \label{fig:gpdt1}
\end{figure}

\begin{figure*}[t]
  \centering
  \includegraphics[width=\textwidth]{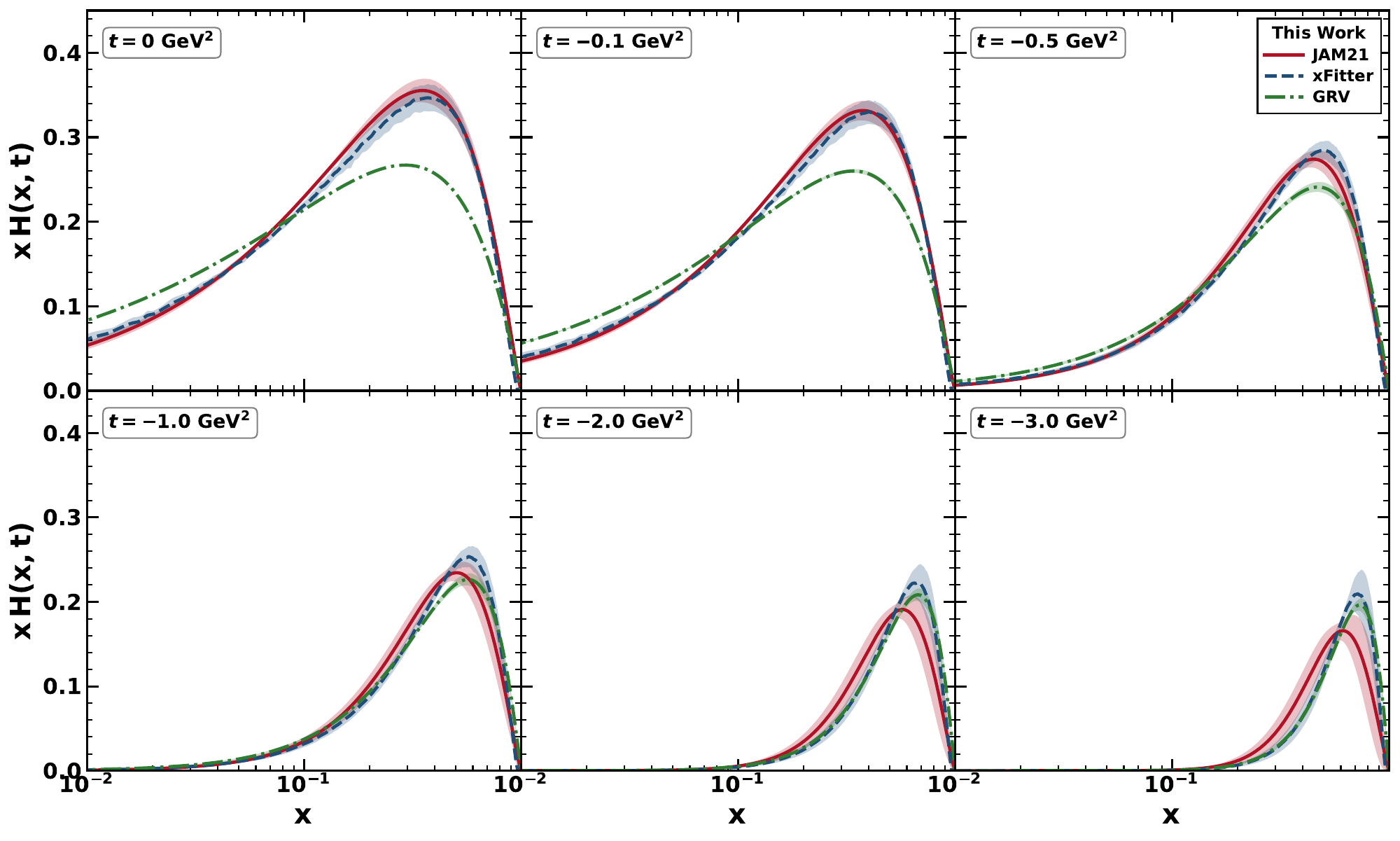}
  \caption{
  Unpolarized valence-quark GPD of the pion from fit~T, shown as
  $x H(x,0,t)$ as a function of $x$ on a logarithmic scale at
  $\mu^2 = 10~\mathrm{GeV}^2$, for $t = 0$, $-0.1$, $-0.5$, $-1.0$, $-2.0$ and
  $-3.0~\mathrm{GeV}^2$. The solid red, dashed blue and dash-dotted green
  curves correspond to the JAM21, xFitter and GRV inputs, and the shaded bands
  are the total uncertainties. The $t=0$ panel reproduces the input
  distributions of Fig.~\ref{fig:pdfs}, since $H(x,0,0) = q_v(x)$.
  }
  \label{fig:gpd}
\end{figure*}

\begin{figure*}[t]
  \centering
  \includegraphics[width=\textwidth]{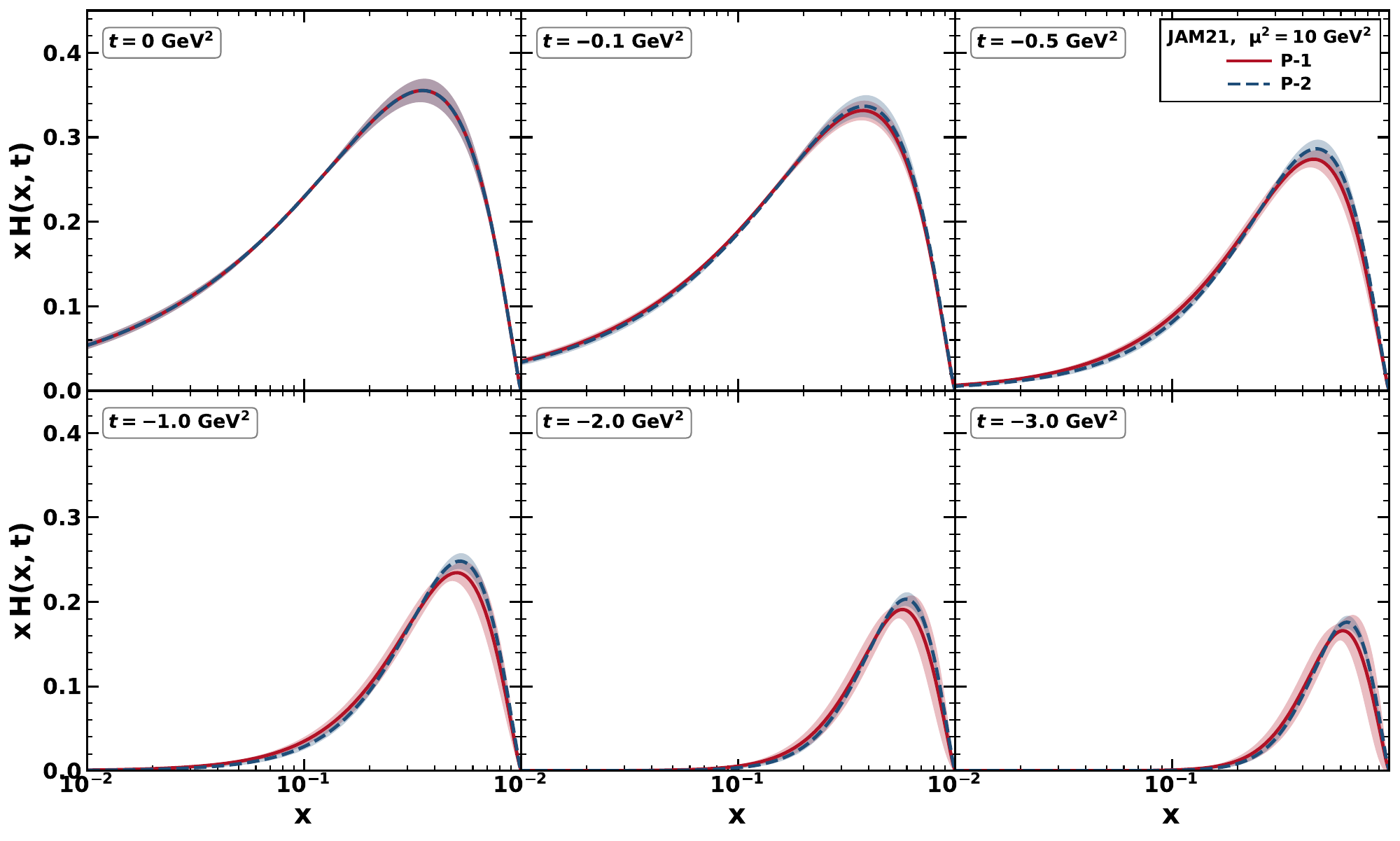}
  \caption{
  Comparison of the two profile functions. The GPD is shown as $x H(x,0,t)$
  as a function of $x$ on a logarithmic scale, from fit~T with the JAM21 input
  at $\mu^2 = 10~\mathrm{GeV}^2$, for $t = 0$, $-0.1$, $-0.5$, $-1.0$, $-2.0$
  and $-3.0~\mathrm{GeV}^2$. The solid red curves are obtained with P-1,
  Eq.~(\ref{eq:P1}), and the dashed blue curves with P-2, Eq.~(\ref{eq:P2}).
  The shaded bands are the total uncertainties. The two profiles coincide at
  $t=0$ and separate progressively as $|t|$ increases.
  }
  \label{fig:cmpGPD}
\end{figure*}

Figure~\ref{fig:gpdt1} shows $xH(x,0,t)$ at $t=-1~\mathrm{GeV}^2$ with the
JAM21 input from all three fits. The peak sits at $x \simeq 0.44$ for fit~E,
at $x \simeq 0.51$ for fit~T and at $x \simeq 0.68$ for fit~L, and the band of
fit~L is about four times wider than that of fit~T, yet each curve reproduces
the form factor it was fitted to. The reason is kinematic rather than
statistical. The experimental points are concentrated below
$|t| \simeq 1~\mathrm{GeV}^2$, where the Regge term of Eq.~(\ref{eq:P1})
dominates and the profile is controlled by its small-$x$ behaviour, whereas
the lattice points lie almost entirely above $|t| \simeq 1.5~\mathrm{GeV}^2$,
where the $\beta x^{m}\ln(1-bt)$ term takes over and the large-$x$ region
carries the integral. Each data set therefore fixes a different corner of the
profile, and the shape at intermediate $x$ follows from extrapolation between
them.

The full $t$ dependence from fit~T is shown in Fig.~\ref{fig:gpd}. At $t=0$
the six panels reproduce the input valence distributions of
Fig.~\ref{fig:pdfs} identically, since Eq.~(\ref{eq:forward}) is an identity
and not a fitted statement. As $|t|$ grows the distribution is suppressed
everywhere, but not uniformly. With the JAM21 input the peak of $xH$ migrates
from $x=0.354$ at $t=0$ to $0.451$ at $t=-0.5~\mathrm{GeV}^2$ and $0.616$ at
$t=-3~\mathrm{GeV}^2$, while its height falls from $0.355$ to $0.166$. The
migration is the statement that large momentum transfers are absorbed
predominantly by fast quarks, the soft-overlap or Feynman mechanism, in which
the recoiling hadron is formed by a single energetic parton that carries
almost the whole momentum while the spectators remain soft. In the present
parameterization this behaviour is enforced by the factor $(1-x)^{\gamma}$,
which switches off the Regge diffusion as $x \rightarrow 1$ and therefore
leaves the large-$x$ region less suppressed at large $|t|$.

\begin{figure}[t]
  \centering
  \includegraphics[width=\columnwidth]{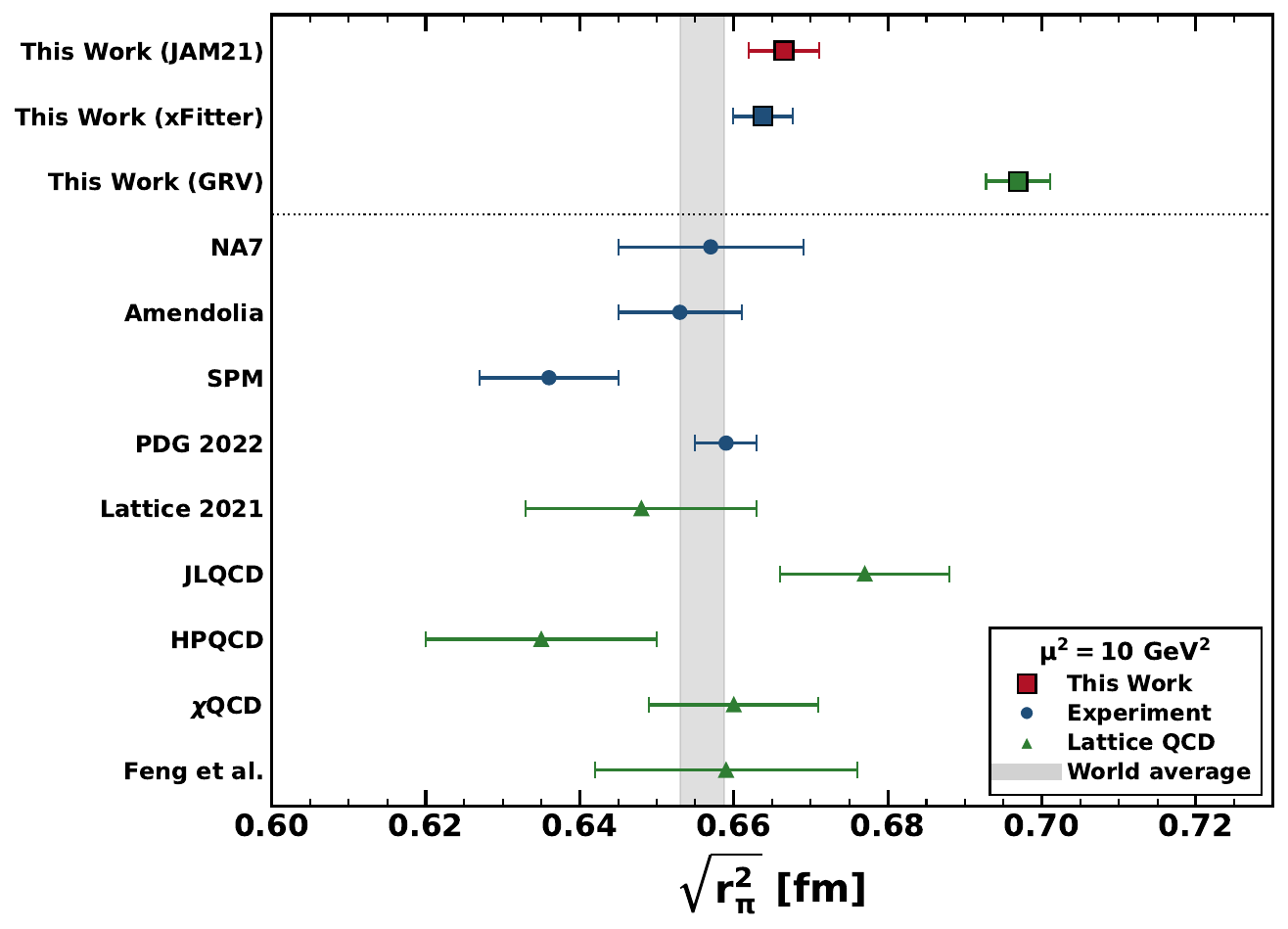}
  \caption{
  Pion charge radius $r_\pi = \sqrt{\langle r_\pi^2\rangle}$ obtained from
  fit~T with the three PDF inputs, compared with the experimental
  determinations of
  Refs.~\cite{Amendolia:1984nz,NA7:1986vav,ParticleDataGroup:2022pth} and the
  lattice-QCD determinations of
  Refs.~\cite{Alexandrou:2021ztx,Cui:2021aee,Aoki:2015pba,Koponen:2015tkr,Wang:2020nbf,Feng:2019geu}.
  The grey band is the weighted average of these nine determinations,
  $0.6559 \pm 0.0028~\mathrm{fm}$.
  }
  \label{fig:radius}
\end{figure}

The pion charge radius follows from the slope of the form factor at the origin,
\begin{align}
\langle r_\pi^2 \rangle
&= -6 \left.\frac{dF(t)}{dt}\right|_{t=0}
\nonumber\\
&= -6 \int_0^1 dx\, q_v(x)
\left.\frac{\partial \mathcal{S}(x,t)}{\partial t}\right|_{t=0}.
\label{eq:r2}
\end{align}
The $x$-dependent transverse width of the quark density is
\begin{equation}
\langle b_\perp^2 \rangle(x)
= \frac{\int d^2\bfb\; b_\perp^2\, q(x,\bfb)}
       {\int d^2\bfb\; q(x,\bfb)}
= 4 \left.\frac{\partial}{\partial t}\ln H(x,0,t)\right|_{t=0},
\label{eq:b2}
\end{equation}
where $q(x,\bfb)$ is the impact-parameter density defined below. Combining
Eqs.~(\ref{eq:r2}) and (\ref{eq:b2}) with the valence normalization gives the
exact model-independent relation
\begin{equation}
\langle b_\perp^2 \rangle
= \int_0^1 dx\, q_v(x)\, \langle b_\perp^2 \rangle(x)
= \frac{2}{3}\,\langle r_\pi^2 \rangle ,
\label{eq:b2r2}
\end{equation}
which we use as a numerical check of the extraction. With the JAM21 input and fit~T we obtain
$\langle r_\pi^2 \rangle = 0.4442~\mathrm{fm}^2$ and
$\langle b_\perp^2 \rangle = 0.2961~\mathrm{fm}^2$, which satisfy
Eq.~(\ref{eq:b2r2}) to four digits. The corresponding charge radius is
$r_\pi = 0.6665 \pm 0.0046~\mathrm{fm}$, and the xFitter input gives
$0.6638 \pm 0.0039~\mathrm{fm}$. Both agree with the weighted world average of
Refs.~\cite{Amendolia:1984nz,NA7:1986vav,ParticleDataGroup:2022pth,Alexandrou:2021ztx,Cui:2021aee,Aoki:2015pba,Koponen:2015tkr,Wang:2020nbf,Feng:2019geu},
$0.6559 \pm 0.0028~\mathrm{fm}$, at the level of two standard deviations or
better, as shown in Fig.~\ref{fig:radius}. Fit~L gives systematically smaller
and much less precise radii, from $0.625$ to $0.651~\mathrm{fm}$ with
uncertainties between $0.025$ and $0.069~\mathrm{fm}$, which is expected since
$\langle r_\pi^2\rangle$ is the slope of $F$ at the origin and the lattice
points begin only at $|t| \simeq 1.5~\mathrm{GeV}^2$.

At zero skewness the GPD is the Fourier conjugate of the impact-parameter
density \cite{Burkardt:2000za},
\begin{equation}
q(x,\bfb)
= \int \frac{d^2\Dp}{(2\pi)^2}\,
e^{-i\,\Dp\cdot\bfb}\,
H(x,0,t=-\Dp^2)
= \rho(x,b_\perp),
\label{eq:fourier}
\end{equation}
which for the axially symmetric case reduces to the Hankel transform
\begin{equation}
\rho(x,b_\perp)
= \frac{1}{2\pi}\int_0^{\infty} d\Delta\,
\Delta\, J_0(b_\perp \Delta)\,
H(x,0,-\Delta^2).
\label{eq:hankel}
\end{equation}
Because the pion is spinless, $\rho(x,\bfb)$ carries no azimuthal dependence
and the transverse image at fixed $x$ is a set of concentric circles. The
$x$-moments of the density are
\begin{equation}
\rho^{\,n}(\bfb)
= \int_0^1 dx\, x^{\,n-1}\, \rho(x,b_\perp),
\label{eq:moments}
\end{equation}
and $n=1$ gives the charge density in the transverse plane.

\begin{figure*}[t]
    \centering
    \includegraphics[width=\textwidth]{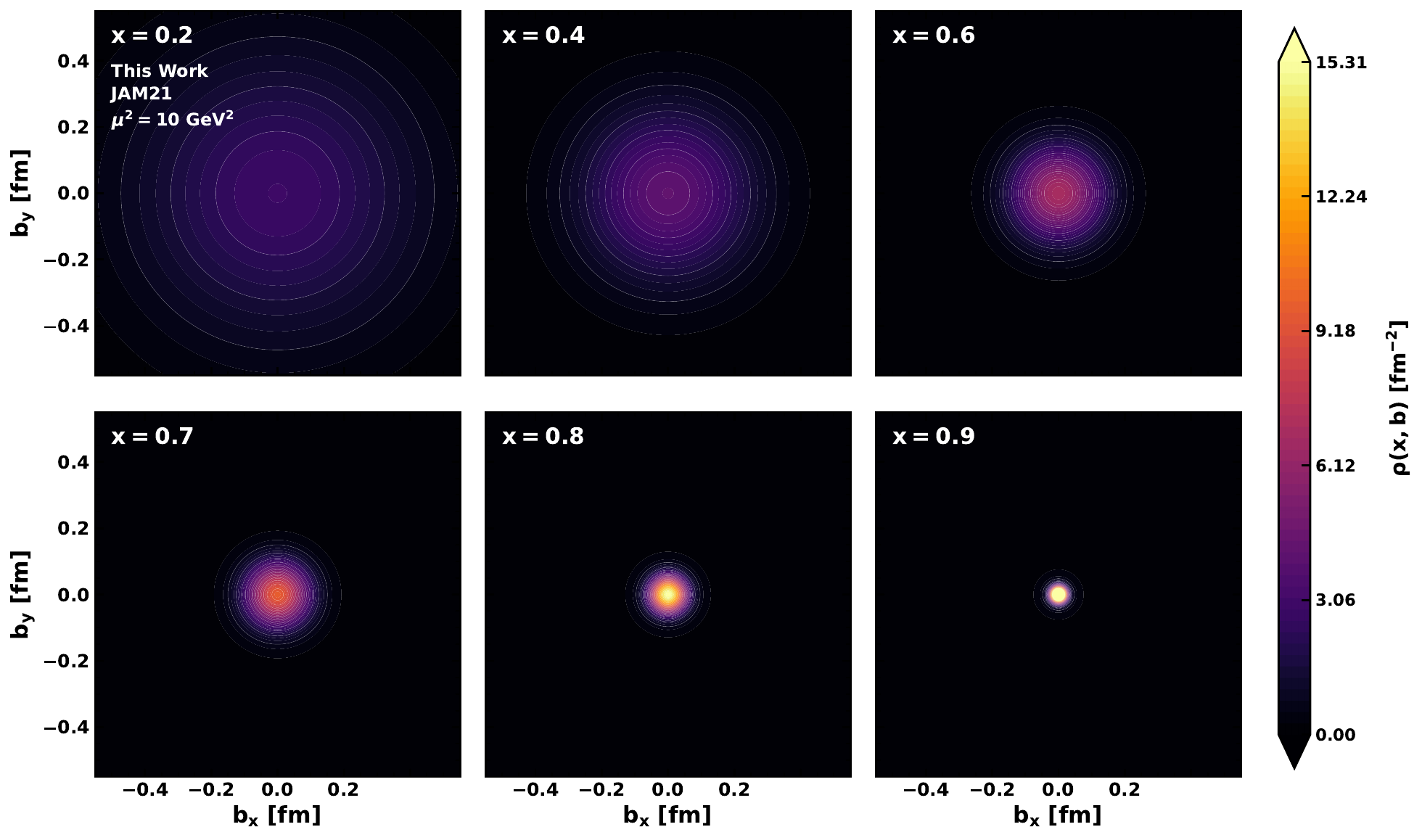}
    \caption{
    Two-dimensional transverse density $\rho(x,\bfb)$ of the pion in the
    $(b_x,b_y)$ plane, from fit~T with the JAM21 input at
    $\mu^2 = 10~\mathrm{GeV}^2$, for $x = 0.2$, $0.4$, $0.6$, $0.7$, $0.8$ and
    $0.9$. The colour bar is linear in $\rho$ and is common to the six panels
    and to Figs.~\ref{fig:tomoxF} and \ref{fig:tomoGRV}, with its upper end
    set at $15.31~\mathrm{fm}^{-2}$, so that the two panels of largest $x$
    saturate it.
    }
    \label{fig:tomoJAM}
\end{figure*}

\begin{figure*}[t]
    \centering
    \includegraphics[width=\textwidth]{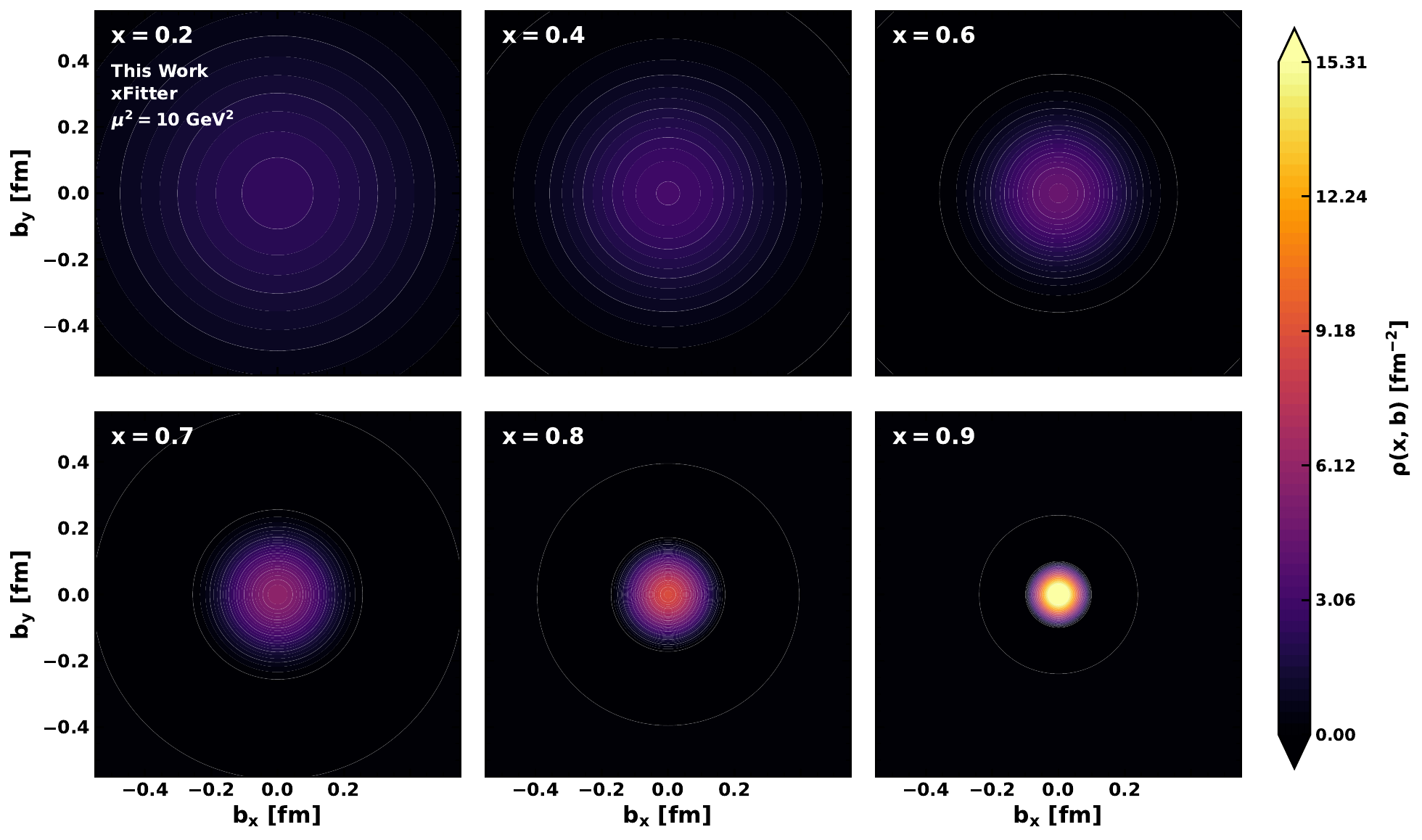}
    \caption{
    Two-dimensional transverse density $\rho(x,\bfb)$ obtained with the
    xFitter input, in the same layout and on the same colour scale as
    Fig.~\ref{fig:tomoJAM}.
    }
    \label{fig:tomoxF}
\end{figure*}

\begin{figure*}[t]
    \centering
    \includegraphics[width=\textwidth]{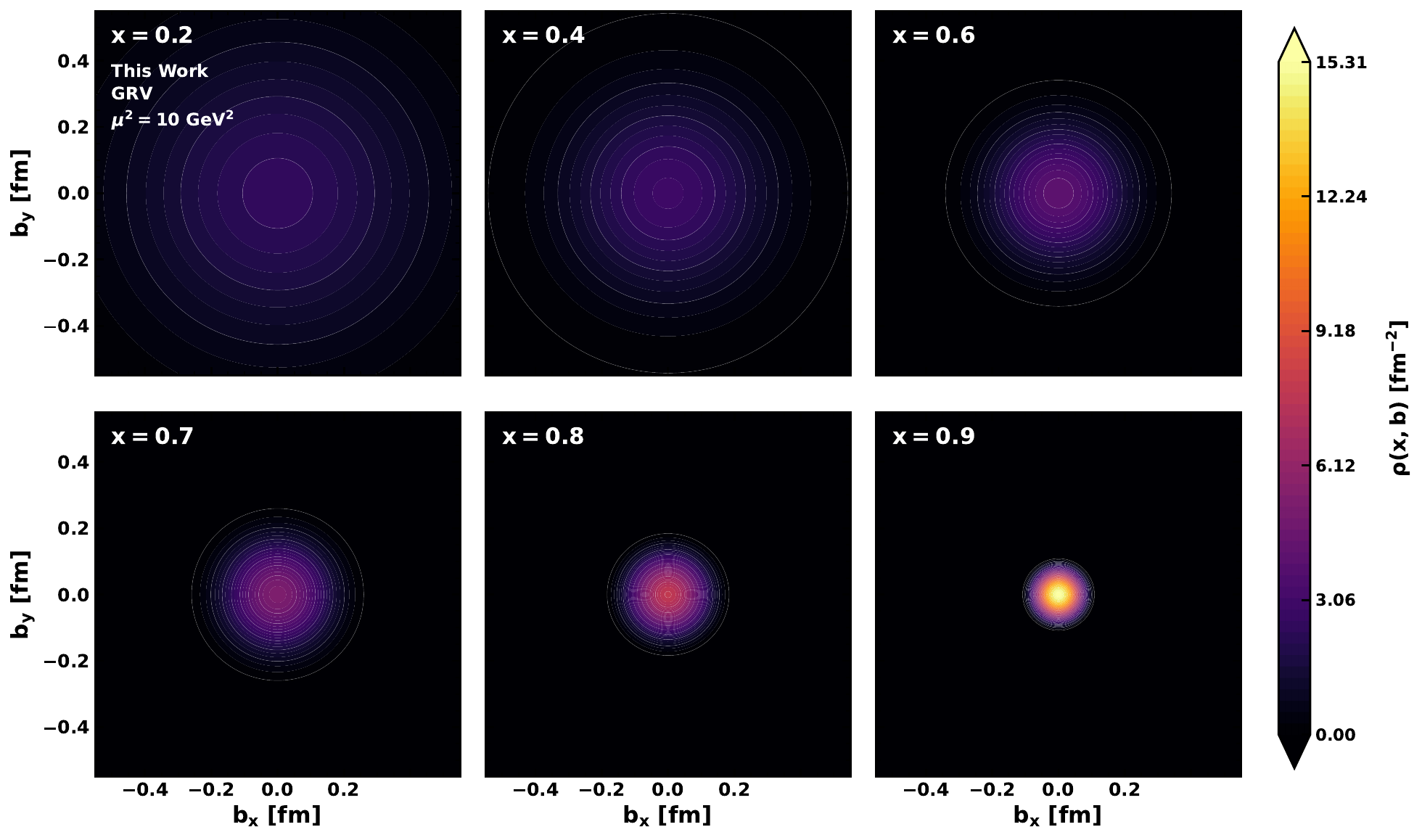}
    \caption{
    Two-dimensional transverse density $\rho(x,\bfb)$ obtained with the GRV
    input, in the same layout and on the same colour scale as
    Fig.~\ref{fig:tomoJAM}.
    }
    \label{fig:tomoGRV}
\end{figure*}

Figures~\ref{fig:tomoJAM}, \ref{fig:tomoxF} and \ref{fig:tomoGRV} show
$\rho(x,\bfb)$ for the three PDF inputs at $x = 0.2$, $0.4$, $0.6$, $0.7$,
$0.8$ and $0.9$. The half-maximum radius falls from about $0.35~\mathrm{fm}$
at $x=0.2$ to about $0.05~\mathrm{fm}$ at $x=0.9$, while the peak height rises
from $2.9$ to $34~\mathrm{fm}^{-2}$ for JAM21, the normalization
$\int d^2\bfb\,\rho(x,\bfb) = q_v(x)$ being preserved throughout. The transverse extent of the pion therefore
decreases rapidly with increasing momentum fraction, the width falling below
half of its value at $x=0.2$ already by $x \simeq 0.5$. The narrowing is required on general grounds, since $\bfb$ is
measured from the transverse centre of momentum and a quark carrying nearly
all the momentum defines that centre itself \cite{Burkardt:2000za}.
Numerically, however, it is produced here by the $(1-x)^{\gamma}$ factor of
Eq.~(\ref{eq:P1}), which the form factor constrains only through its $x$
integral, so the rate of the contraction is a prediction of the
parameterization rather than a measurement.

\begin{figure}[t]
    \centering
    \includegraphics[width=\columnwidth]{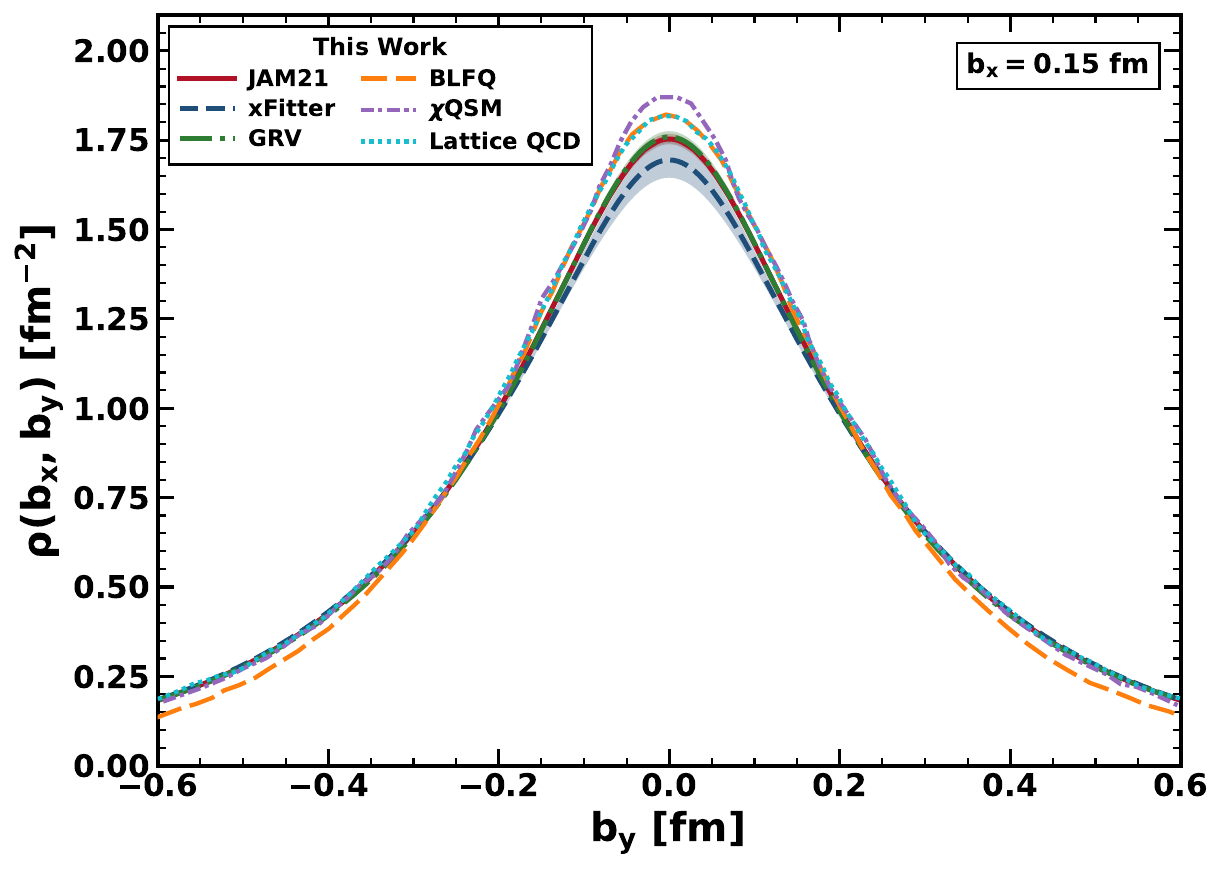}
    \caption{
    Transverse density $\rho(b_x,b_y)$ of the pion, Eq.~(\ref{eq:moments})
    with $n = 1$, as a function of $b_y$ at fixed $b_x = 0.15~\mathrm{fm}$ and
    $\mu^2 = 10~\mathrm{GeV}^2$, from fit~T with the three PDF inputs. The
    shaded bands are the total uncertainties of Sec.~2. Also shown are the
    BLFQ result of Ref.~\cite{Adhikari:2021jrh}, the chiral quark-soliton
    model of Ref.~\cite{Nam:2010pt} and the lattice-QCD calculation of
    Ref.~\cite{QCDSF:2007ifr}, none of which entered the fit.
    }
    \label{fig:density}
\end{figure}

Figure~\ref{fig:density} shows the $n=1$ moment $\rho(b_x,b_y)$ at fixed
$b_x = 0.15~\mathrm{fm}$ as a function of $b_y$, together with the BLFQ
calculation of Ref.~\cite{Adhikari:2021jrh}, the chiral quark-soliton model of
Ref.~\cite{Nam:2010pt} and the lattice-QCD result of
Ref.~\cite{QCDSF:2007ifr}. Our peak densities, $1.75~\mathrm{fm}^{-2}$ for
JAM21, $1.69~\mathrm{fm}^{-2}$ for xFitter and $1.77~\mathrm{fm}^{-2}$ for
GRV, bracket the lattice and model values, and the shapes agree over the whole
range $|b_y| \le 0.6~\mathrm{fm}$. None of these calculations entered the
fit, and the three proceed from different dynamical descriptions, a
light-front Hamiltonian, a soliton in the chiral limit and a direct simulation
of the Euclidean correlator. Their agreement on a common transverse charge
profile indicates that this observable is governed largely by the valence
normalization and the charge radius, and is therefore insensitive to the
details of the underlying model.

We note that the present analysis is restricted to the unpolarized vector GPD.
The transverse spin density of a quark with transverse polarization
$\vec{s}_\perp$ inside the pion involves in addition the chiral-odd GPD
$H_T^q$ through
\begin{equation}
\rho(x,\bfb,\vec{s}_\perp)
= \frac{1}{2}\left[
q(x,\bfb)
- \frac{s_\perp^i \epsilon^{ij}_\perp b_\perp^j}{M_\pi}
\frac{\partial}{\partial b_\perp^2}\, q_T(x,\bfb)
\right],
\label{eq:spindensity}
\end{equation}
where $q_T$ is the impact-parameter transform of $H_T^q(x,0,t)$. The tensor
GPD is constrained by the pion tensor form factor, for which no experimental
measurement exists and only limited lattice information is available
\cite{QCDSF:2007ifr,Alexandrou:2021ztx}. We therefore set the second term of
Eq.~(\ref{eq:spindensity}) aside and quote only the unpolarized density.
Transverse spin densities in the pion have been examined in
Refs.~\cite{Adhikari:2021jrh,QCDSF:2007ifr,Nam:2010pt,Diehl:2005jf,Pasquini:2007xz,Fanelli:2016aqc}.

The two profiles differ in one structural respect. In P-2 the exponent is
linear in $t$, so that $H$ falls exponentially with $|t|$ at fixed $x$,
whereas the $\ln(1-bt)$ term of P-1 grows only logarithmically and produces a
power-law falloff. Both reduce to $q_v(x)$ at $t=0$ and both carry the same
Regge behaviour at small $x$, so the difference is confined to large $|t|$ and
moderate to large $x$.

\begin{figure*}[t]
  \centering
  \includegraphics[width=\textwidth]{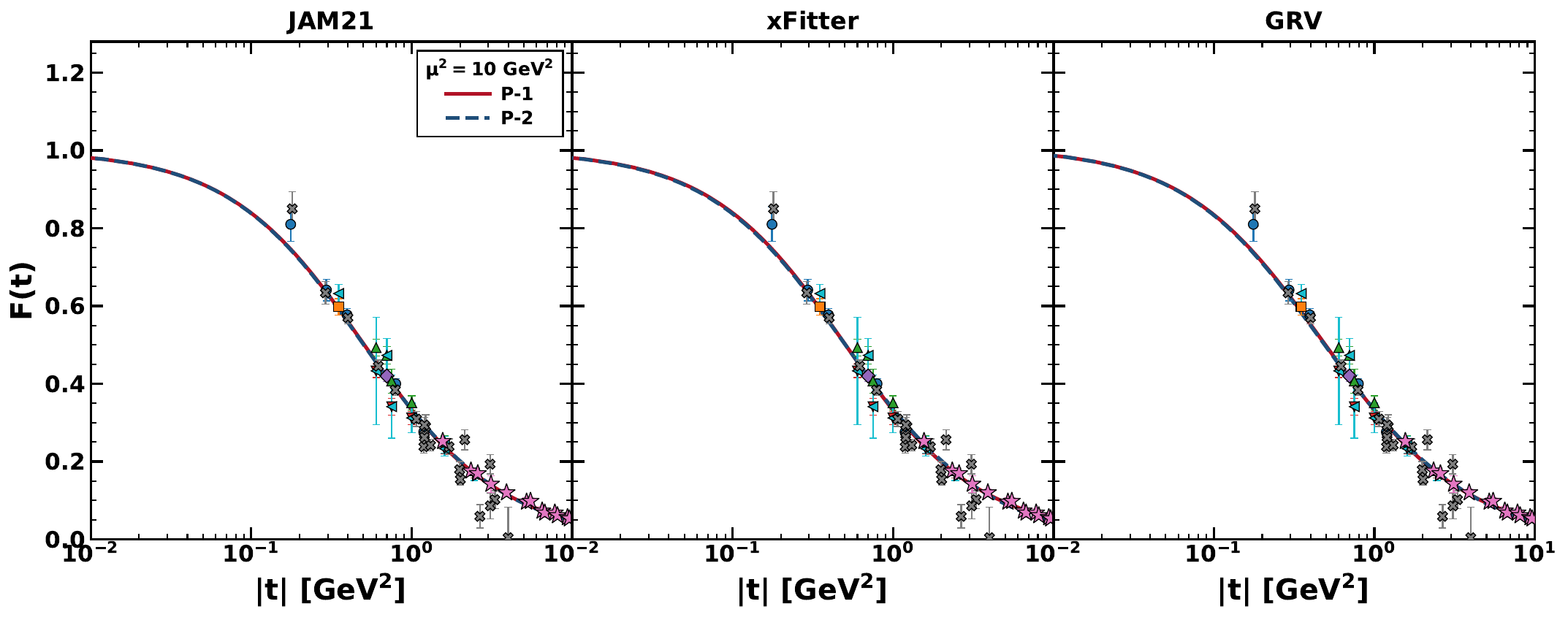}
  \caption{
  Pion electromagnetic form factor $F(t)$ from fit~T with the two profile
  functions, shown for the JAM21, xFitter and GRV inputs from left to right at
  $\mu^2 = 10~\mathrm{GeV}^2$. The solid red curves are obtained with P-1,
  Eq.~(\ref{eq:P1}), and the dashed blue curves with P-2, Eq.~(\ref{eq:P2}).
  The experimental and lattice-QCD data are as in Fig.~\ref{fig:fitT}.
  }
  \label{fig:cmpF}
\end{figure*}

\begin{figure*}[t]
  \centering
  \includegraphics[width=\textwidth]{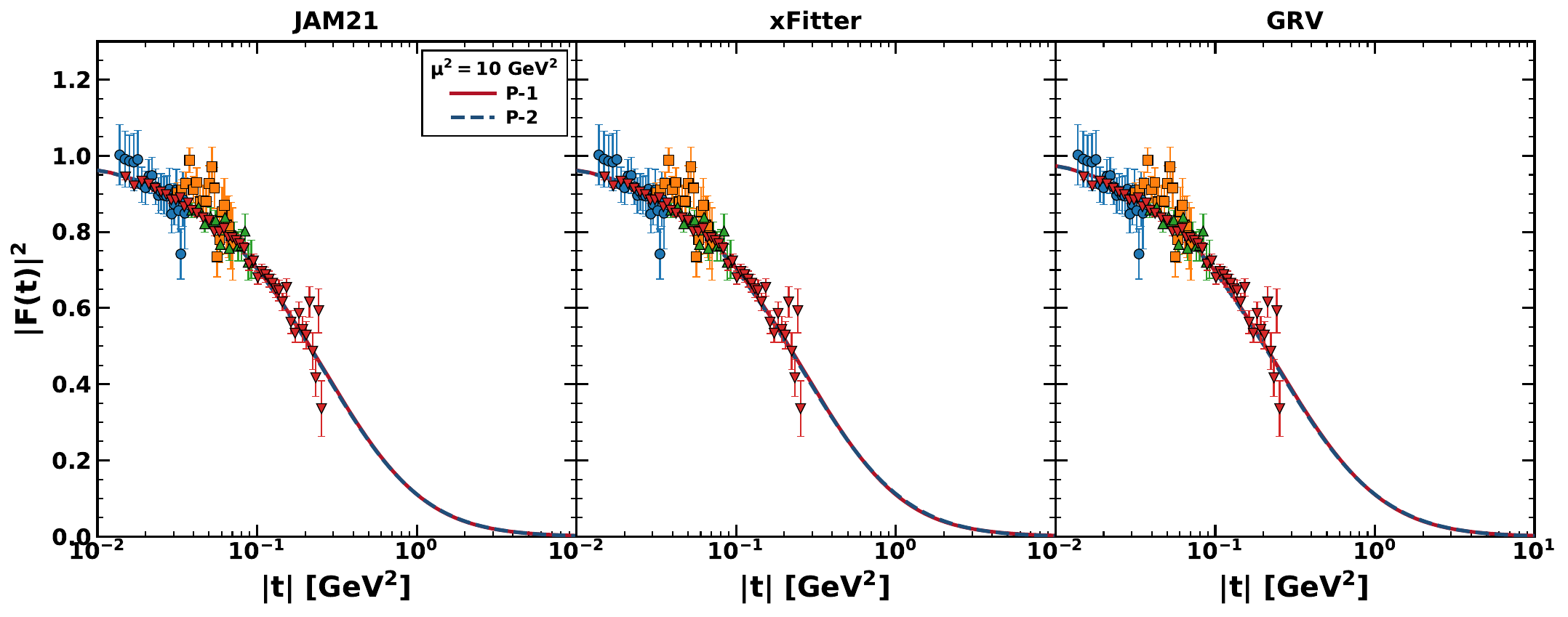}
  \caption{
  Squared pion electromagnetic form factor $|F(t)|^2$ from fit~T with the two
  profile functions, in the same layout and with the same curve convention as
  Fig.~\ref{fig:cmpF}. The experimental data are taken from
  Refs.~\cite{Adylov:1977kj,Dally:1981ur,Dally:1982zk,NA7:1986vav}.
  }
  \label{fig:cmpF2}
\end{figure*}

This difference is not resolved by the form factor. Figures~\ref{fig:cmpF}
and \ref{fig:cmpF2} show $F(t)$ and $|F(t)|^2$ from the two profiles, and the
curves lie on top of one another over the whole measured range. In the
combined fit P-1 gives $\chi^2/N = 1.45$, $1.44$ and $1.72$ for JAM21,
xFitter and GRV, compared with $1.48$, $1.51$ and $1.78$ for P-2, so the two
agree to within a few per cent for every input even though P-1 carries five
parameters and P-2 only three. The charge radii of Table~\ref{tab:obs} agree
still more closely, by
$0.0002~\mathrm{fm}$ for JAM21 in the combined fit and by at most
$0.006~\mathrm{fm}$ across the three inputs. Any quantity obtained from an $x$
integral of the GPD is thus common to the two.

The GPD itself is not common to the two. Figure~\ref{fig:cmpGPD} compares
$xH(x,0,t)$ from the two profiles with the JAM21 input. The curves coincide at $t=0$ and stay
within their bands up to $|t| \simeq 0.1~\mathrm{GeV}^2$, then separate
progressively, with P-2 above P-1 near the peak, until at
$t=-3~\mathrm{GeV}^2$ they differ by more than the width of either band over
an appreciable range of $x$. The separation grows precisely where the two
functional forms diverge, which identifies the falloff in $t$ at fixed $x$ as
the feature that the present data leave open. Discriminating between a
power-law and an exponential falloff requires an observable differential in
$x$, either the exclusive reactions listed in Sec.~1 or a lattice
determination of the light-cone GPD, and such a measurement would provide the
strongest improvement to an extraction of this kind.

The same conclusion follows from the lattice-only fits. With P-2 these return
charge radii between $0.669$ and $0.718~\mathrm{fm}$, compared with $0.625$
to $0.651~\mathrm{fm}$ for P-1, a spread considerably larger than the
corresponding one in the combined fits. Thirteen points confined to large
$|t|$, with uncertainties comparable to their scatter, do not determine the
slope of $F$ at the origin, which is then fixed largely by the assumed shape
of the profile.

{\it\textbf{4. Summary.}}
We have determined the unpolarized valence-quark generalized parton
distribution of the pion at zero skewness, $H(x,0,t)$, from the combined
experimental and lattice-QCD information on the pion electromagnetic form
factor, using three global pion PDF sets as input without any intermediate
refit. Three separate fits, to the experimental data alone, to the lattice
data alone and to the two together, were performed and compared. The
experimental and lattice form-factor data are mutually consistent, in the
sense that adding the lattice points leaves the description of the
experimental points unchanged, although the large lattice uncertainties limit
the strength of that statement. The uncertainty of the extraction combines the
$\Delta\chi^2=1$ region of the profile parameters, sampled rather than
propagated linearly because several parameters lie at a boundary, with the
propagated uncertainty of the input PDF error members.

The charge radius obtained from the combined fit with the profile P-1 is
$r_\pi = 0.6665 \pm 0.0046~\mathrm{fm}$ with the JAM21 input and
$0.6638 \pm 0.0039~\mathrm{fm}$ with xFitter, both close to the world average
$0.6559 \pm 0.0028~\mathrm{fm}$, while the GRV input gives
$0.6969 \pm 0.0042~\mathrm{fm}$ and a total $\chi^2/N$ larger by about $0.28$,
so that this set is disfavoured by the form-factor data. The extracted
transverse width falls monotonically with $x$, and the two-dimensional density
contracts towards the transverse centre of the pion as $x$ grows, in agreement
with BLFQ, chiral quark-soliton and lattice-QCD calculations that were not
used in the fit.

The principal limitation of the approach is set by the observable itself. The
form factor is an $x$ integral, so the $x$ dependence of $H(x,0,t)$ at fixed
$t$ follows from the assumed profile function rather than from the data. We
have quantified this by repeating the analysis with the profile P-2 used
previously by the MMGPDs group, which describes the form factor essentially as
well as P-1 and gives charge radii agreeing to better than
$0.006~\mathrm{fm}$, yet differs from P-1 in $H(x,0,t)$ by more than the
quoted uncertainty bands once $|t|$ exceeds about $1~\mathrm{GeV}^2$. Direct
measurements of exclusive processes at the 12~GeV Jefferson Lab program, at
COMPASS and AMBER, and at the future Electron-Ion Collider through the
Sullivan process, together with lattice-QCD determinations of the light-cone
GPD, are needed to reduce this residual model dependence.

{\it\textbf{5. Acknowledgements.}}
S.P.\ thanks Prof.\ Oleg V.\ Teryaev for fruitful discussions. He gratefully
acknowledges the Bogoliubov Laboratory of Theoretical Physics (BLTP), Joint
Institute for Nuclear Research (JINR), Dubna, for providing research
facilities and support during his visit, where this work was carried out.
H.D.\ acknowledges financial support from the Science and Engineering Research
Board (SERB), Anusandhan National Research Foundation, Government of India,
under the SERB-POWER Fellowship (Ref.\ No.\ SPF/2023/000116).

\bibliography{ref_pion.bib}

\section{Appendix}

\end{document}